\documentclass[AMA,Times2COL]{WileyNJDv5}

\usepackage{siunitx} 
\makeatletter
\newcommand*{\addFileDependency}[1]{%
	\typeout{(#1)}%
	\@addtofilelist{#1}%
	\IfFileExists{#1}{}{\typeout{No file #1.}}%
}
\makeatother

\usepackage{siunitx}
\usepackage{hyperref}
\hypersetup{
	colorlinks=true,
	urlcolor=blue
}
\usepackage{xr}

\newcommand*{\myexternaldocument}[1]{%
	\externaldocument{#1}%
	\addFileDependency{#1.tex}%
	\addFileDependency{#1.aux}%
}

\makeatletter
\@ifundefined{r@sec:supp_sample}{\newlabel{sec:supp_sample}{{S1}{1}}}{}
\@ifundefined{r@sec:supp_setup}{\newlabel{sec:supp_setup}{{S2}{1}}}{}
\@ifundefined{r@sec:supp_binary_derivation}{\newlabel{sec:supp_binary_derivation}{{S3}{1}}}{}
\@ifundefined{r@sec:supp_field_optimizations}{\newlabel{sec:supp_field_optimizations}{{S4.1}{1}}}{}
\@ifundefined{r@sec:supp_SLM_optimizations}{\newlabel{sec:supp_SLM_optimizations}{{S4.2}{1}}}{}
\@ifundefined{r@sec:supp_Index_match}{\newlabel{sec:supp_Index_match}{{S5}{1}}}{}
\@ifundefined{r@sec:supp_incSIM}{\newlabel{sec:supp_incSIM}{{S6}{1}}}{}
\@ifundefined{r@sec:supp_powerSpectrum}{\newlabel{sec:supp_powerSpectrum}{{S7}{1}}}{}
\@ifundefined{r@sec:supp_FullFOV}{\newlabel{sec:supp_FullFOV}{{S8}{1}}}{}
\makeatother

\myexternaldocument{sm1_HC}

\articletype{Research Article}%

\begin{document}
	
	\title{Tailored Speckle Illumination Microscopy with Enhanced Sectioning and Image Quality}
	
	\author[1]{SeungYun Han}
	\author[1]{KyeoReh Lee}
	\author[2]{Young Seo Kim}
	\author[3,4]{Chuan Li}
	\author[5]{Nicholas Bender}
	\author[6]{Kabish Wisal}
	\author[2]{Taeyun Ku}
	\author[4,7]{Jerome Mertz}
	\author[1]{Hui Cao}
	
	
	
	\address[1]{\orgdiv{Department of Applied Physics}, \orgname{Yale University}, \orgaddress{\state{New Haven, CT 06510}, \country{USA}}}
	
	\address[2]{\orgdiv{Graduate School of Medical Science and Engineering}, \orgname{Korea Advanced Institute of Science and Technology (KAIST)}, \orgaddress{\state{Daejeon 34141}, \country{Republic of Korea}}}
	
	\address[3]{\orgdiv{Department of Electrical \& Computer Engineering}, \orgname{Boston University}, \orgaddress{\state{Boston, MA 02215}, \country{USA}}}
	
	\address[4]{\orgdiv{Photonics Center}, \orgname{Boston University}, \orgaddress{\state{Boston, MA 02215}, \country{USA}}}
	
	\address[5]{\orgdiv{School of Applied \& Engineering Physics}, \orgname{Cornell University}, \orgaddress{\state{Ithaca, NY 14853}, \country{USA}}}
	
	\address[6]{\orgdiv{Department of Physics}, \orgname{Yale University}, \orgaddress{\state{New Haven, CT 06510}, \country{USA}}}
	
	\address[7]{\orgdiv{Department of Biomedical Engineering}, \orgname{Boston University}, \orgaddress{\state{Boston, MA 02215}, \country{USA}}}
	
	\corres{Corresponding author Hui Cao, \email{hui.cao@yale.edu}}
	
	
	
	\abstract[Abstract]{Optical speckle patterns have been widely used for illumination in computational imaging, optical sectioning microscopy, and super-resolution imaging. However, commonly used speckles satisfy Rayleigh statistics, which are not ideal for diverse imaging applications. Here we tailor three-dimensional speckle intensity statistics for dynamic speckle illumination microscopy based on linear fluorescence. Optical sectioning is enhanced by axially varying speckle contrast, and image reconstruction noise is minimized with in-focus speckles of binary intensities. The customized speckle statistics are shown to tolerate sample-induced aberration and scattering. We apply tailored speckle illumination to mouse brain vascular imaging and demonstrate much improved image quality than optical-sectioning structured illumination. These results establish customization of speckle intensity statistics as a promising strategy for robust, high-throughput fluorescence imaging in thick, scattering biological specimens.}
	
	
	\jnlcitation{\cname{%
			\author{},
			\ctitle{} \cjournal{\it} \cvol}
	}
	
	\maketitle
	
	
		\section{Introduction}
		
		Optical speckles are granular intensity fluctuations resulting from interference of scattered waves with varying phase. 
		Over the past two decades, speckle patterns have been widely used for illumination in ghost imaging \cite{zerom2012thermal, zhang2015ghost, mao2016speckle, kuplicki2016high, leibov2021speckle, zhang2022optimizing, yang2023underwater} and compressive imaging \cite{bernet2011lensless, bertolotti2012non, almoro2012enhanced, harm2014lensless, katz2014non, lee2016exploiting,  antipa2017diffusercam,  salhov2018depth, yeh2019speckle, yoon2020deep, cao2022dynamic, kang2023mapping,  premillieu2024aberration}.
		High spatial frequencies and optical vortices in speckle patterns enable super-resolution imaging \cite{mudry2012structured, kim2015superresolution, yilmaz2015speckle, zhang2016high, idier2017superresolution, yeh2017structured, mangeat2021super, bender2021circumventing, mazzella2024extended, xing2025spatiotemporal}.   	Axially rapid, non-periodic changes of speckle patterns provide optical sectioning \cite{lim2008wide, lim2011optically, mazzaferri2011analyzing, schniete2018fast, pascucci2019compressive, zheng2023multiplane}.
        Commonly used speckles satisfy Rayleigh statistics and have an intensity contrast of unity, thus called Rayleigh speckles \cite{goodman1975statistical, goodman2007speckle}. 
		Sample-induced aberration and scattering do not modify Rayleigh statistics, making speckle-illumination methods tolerant to wavefront distortions. 
		However, Rayleigh speckles are not necessarily ideal illumination patterns for different applications. 
		
		Our goal is to customize speckle intensity statistics for specific imaging modalities. Here we focus on dynamic speckle illumination (DSI) microscopy, which uses three-dimensional (3D) speckle illumination to achieve optically sectioned fluorescence imaging of thick, inhomogeneous specimens \cite{ventalon2005quasi, ventalon2006dynamic, mertz2011optical}. Illuminating a fluorescent sample sequentially with different Rayleigh speckles, fluctuations in the recorded fluorescence intensity provide an in-focus image, without raster scanning as in confocal microscopy. However, DSI requires a large number of images with different speckle illumination (frames) to reduce reconstruction noise. This not only limits the imaging speed but also increases the likelihood of sample photobleaching and image smearing by sample drift. While HiLo microscopy needs only one image with speckled illumination and another one with uniform illumination, it still suffers from residual speckle noise and has limited background rejection \cite{lim2011optically, mazzaferri2011analyzing, michaelson2012depth, lauterbach2015fast, shi2021evaluating, hu2024hilo}. Moreover, DSI sectioning capability with Rayleigh speckles is typically weaker than that of optical-sectioning structured illumination microscopy (OS-SIM) based on periodic fringes \cite{neil1997method}.
		
		In this work we tailor 3D speckle statistics to simultaneously enhance DSI axial sectioning and reduce the number of frames for image reconstruction. To decrease reconstruction noise from finite sampling, we customize the in-focus speckle pattern of binary intensity values, and reduce the number of frames by threefold from Rayleigh speckles. Axial sectioning strength is further enhanced by lowering the intensity contrast of speckles out of focus. We experimentally demonstrate that tailored speckle illumination remains effective under sample-induced aberration and scattering. Finally we apply tailored speckle illumination to vascular imaging of mouse brain and obtain a signal-to-noise ratio (SNR) significantly higher than that with Rayleigh speckle and OS-SIM.
		
		
		\begin{figure*}[htbp]
			\centering
			\includegraphics[width=0.9\textwidth,
			trim=185pt 10pt 120pt 12pt,clip]{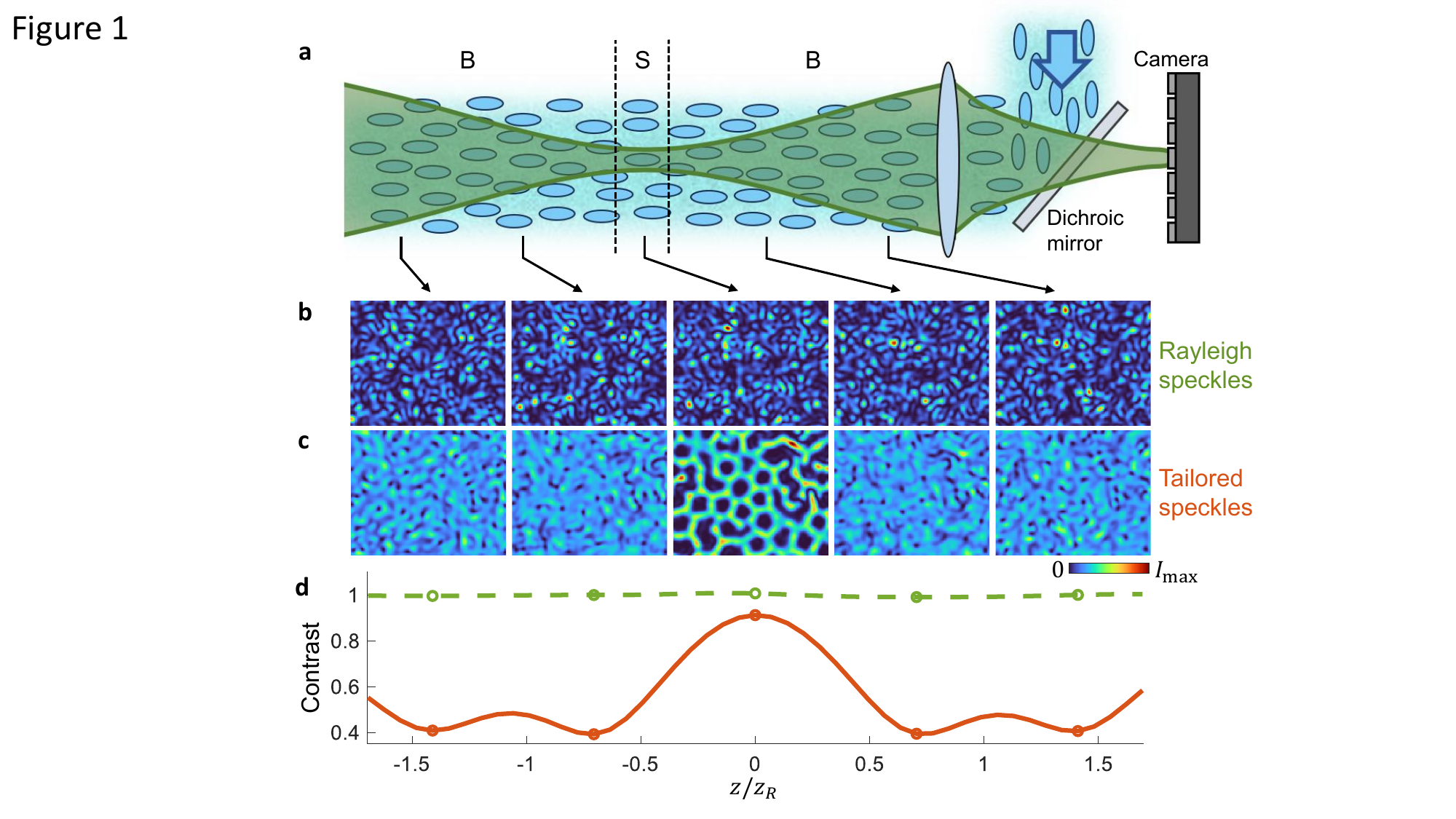} 
			\caption{\textbf{3D speckles with tailored statistics.}
			(a) Schematic of epi-fluorescence imaging with speckle illumination. A thick fluorescent sample is excited by 3D speckled light. A single camera pixel records all fluorescence within the green shaded region, including both in-focus S and out-of-focus B. (b,c) Examples of excitation intensity patterns at different axial positions for (b) Rayleigh speckles and (c) tailored speckles. (d) Speckle intensity contrast as a function of normalized axial positions $z/z_R$. Rayleigh speckles (green dashed) maintain unity contrast axially, but tailored speckles (orange solid) exhibit high contrast at focus and low contrast away from focus.} \label{fig_schematic}
		\end{figure*}
		
		\section{Speckle customization}
		
		Figure~\ref{fig_schematic}a is a schematic of DSI microscopy in epi geometry. A thick fluorescent sample is illuminated by 3D speckles of excitation light (blue). Its linear fluorescence is collected by a lens and projected to a camera. All fluorescence within the green shaded region reaches one camera pixel. It contains fluorescence from the in-focus plane and outside, denoted as S and B, respectively. Since the out-of-focus region B is much larger than the in-focus S, the background fluorescence $I^B$ from B overwhelms the signal $I^S$ from S, $I^B \gg I^S$. Consequently the total fluorescence $I = I^S + I^B$ is dictated by $I^B$, $I \approx I^B$. 
		
		However, when illuminating the sample sequentially with uncorrelated speckle patterns, the fluctuation of $I^S$ from frame to frame is much stronger than that of $I^B$. This is because region B contains many more speckle grains than S, and summing fluorescence from these grains reduces the fluctuation of $I^B$. Thus, standard deviation of $I^S$ is much larger than that of $I^B$ and is approximately equal to that of $I$, $\sigma_N[I] \approx \sigma_N[I^S] \gg \sigma_N[I^B]$, where   $\sigma_N(\cdot)$ denotes the standard deviation over $N$ illumination patterns.  
		
		For simplicity we ignore the detection point spread function (PSF), and write $I^S_n(\mathbf{r})~=~M_n(\mathbf{r})\,O(\mathbf{r})$,
		where $\mathbf{r}$ denotes the lateral position over the in-focus plane, $n = 1, 2, ..., N$, $I^S_n(\mathbf{r})$ and $M_n(\mathbf{r})$ are the $n$-th in-focus fluorescence and illumination intensity patterns respectively, and $O(\mathbf{r})$ denotes the fluorescence source distribution. In a static (time-invariant) sample, $\sigma_N[I^S_n(\mathbf{r})] = \sigma_N[M_n(\mathbf{r})] \, O(\mathbf{r})$. If $N$ is sufficiently large, $\sigma_N[M_n(\mathbf{r})]$ becomes approximately uniform in space, and $\sigma_N[I^S_n(\mathbf{r})]$ is proportional to $O(\mathbf{r})$. Since $\sigma_N[I^S_n(\mathbf{r})] \approx \sigma_N[I_n(\mathbf{r})]$, the in-focus object can be reconstructed from:
		\begin{equation}\label{Eq:DSIrecon}
			I_{\mathrm{recon}}(\mathbf{r}) = \sigma_N[I_n(\mathbf{r})]
			\approx \sigma_N[I^S_n(\mathbf{r})]
			= \sigma_N [M_n(\mathbf{r})]\, O(\mathbf{r}) \propto O(\mathbf{r}) \, .
		\end{equation}
		Thus, DSI provides the in-plane image with a depth discrimination of out-of-focus objects. 
		
		\subsection{Speckle noise suppression} \label{sec:BinaryDerivation}
		Reconstruction noise with DSI is from the spatial inhomogeneity of $\sigma_N[M_n(\mathbf r)]$ due to the finite number of frames. We therefore seek the speckle intensity statistics of excitation light to minimize spatial variance of $\sigma_N[M_n(\mathbf r)]$ for a given $N$. Below we assume (i) every illumination speckle pattern $M_n(\mathbf r)$ is independently sampled from the same intensity probability density function (PDF), (ii) each pattern contains sufficiently many speckle grains such that mean intensity $\langle M_n(\mathbf r) \rangle_r$ and standard deviation $\sigma_{\mathbf r}(M_n(\mathbf r))$ over $\mathbf r$ is identical for different $n$, where $\langle\cdot\rangle_{\mathbf r}$ and $\sigma_{\mathbf{r}}(\cdot)$ denotes the average and standard deviation over spatial coordinate $\mathbf r$, respectively. 
		While minimizing the spatial fluctuation of $\sigma_N[M_n(\mathbf r)]$, its own magnitude should not be lowered, otherwise it would compromise the DSI signal. Therefore, to compare spatial fluctuation with a fixed mean and variance of illumination patterns, we standardize $M_n(\mathbf r)$ as
		\begin{equation}\label{eq:normalize}
			J_n(\mathbf r)=\frac{
				M_n(\mathbf r)-\left\langle M_n(\mathbf 
				r)\right\rangle_{\mathbf r}
			}
			{
				\sigma_{\mathbf{r}}\!\left( M_n(\mathbf r)\right)
			} \, ,
		\end{equation}
		 so that $\langle J_n(\mathbf r) \rangle_{\mathbf r} = 0$, and $\sigma_{\mathbf{r}}(J_n(\mathbf r)) = 1$. 
		 
		 Since $\sigma_N(J_n(\mathbf r)) \ge 0$, making $\sigma_N(J_n(\mathbf r))$ spatially uniform is equivalent to requiring its squared value to be spatially uniform. We therefore define the frame variance
		 	$X(\mathbf r)\equiv \left\{\sigma_N[J_n(\mathbf r)]\right\}^2=\mathrm{Var}_N\left[J_n(\mathbf r)\right]$, 
		 where $\mathrm{Var}_N(\cdot)$ denotes the variance over $N$ illumination patterns (frames). We quantify the spatial non-uniformity of $X(\mathbf r)$ by its spatial contrast,
		 	$C_X \equiv {\sigma_{\mathbf r}[X(\mathbf r)]} / {\langle X(\mathbf r)\rangle_{\mathbf r}}$.		
		 As detailed in Sec.~\ref{sec:supp_binary_derivation} of the Supplementary Materials, we derive 
		 \begin{equation}\label{eq:calculationResult}
		 	C_X^2=
		 	\frac{1}{N} \mathrm{Var}_{\mathbf r} \left(J_n^{2}(\mathbf r)\right)
		 	+\frac{2}{N(N-1)} \, , 
		 \end{equation} 
		 where $\mathrm{Var}_{\mathbf r}(\cdot)$ denotes the spatial variance over ${\mathbf r}$. 
		 
		 Equation~\ref{eq:calculationResult} implies that minimizing $C_X^2$ for a given $N$ is to set $\mathrm{Var}_{\mathbf r}\left(J_n^{2}(\mathbf r)\right) = 0$. The vanishing variance implies constant $J_n^{2}(\mathbf r)$ in space, which corresponds to $J_n^{2}(\mathbf r) = 1$ as the standardization in Eq.~\ref{eq:normalize} sets $\langle J_n^{2}(\mathbf r) \rangle_{\mathbf r}= 1$. Together with the constraint $\langle J_n(\mathbf r) \rangle_{\mathbf r} = 0$ imposed by the standardization, the unique optimal PDF $P(J)$ is
		\begin{equation}\label{Binary PDF}
			P\left(J\right)=
			\begin{cases}
				1/2, & J=\pm 1,\\
				0, & \text{otherwise},
			\end{cases}
		\end{equation}
		which is a \emph{binary} distribution. The corresponding speckle patterns have binary intensity values.  
		
		\begin{figure*}[b]
			\centering
			\includegraphics[width=0.9\textwidth,
			trim=90pt 105pt 110pt 92pt,clip]{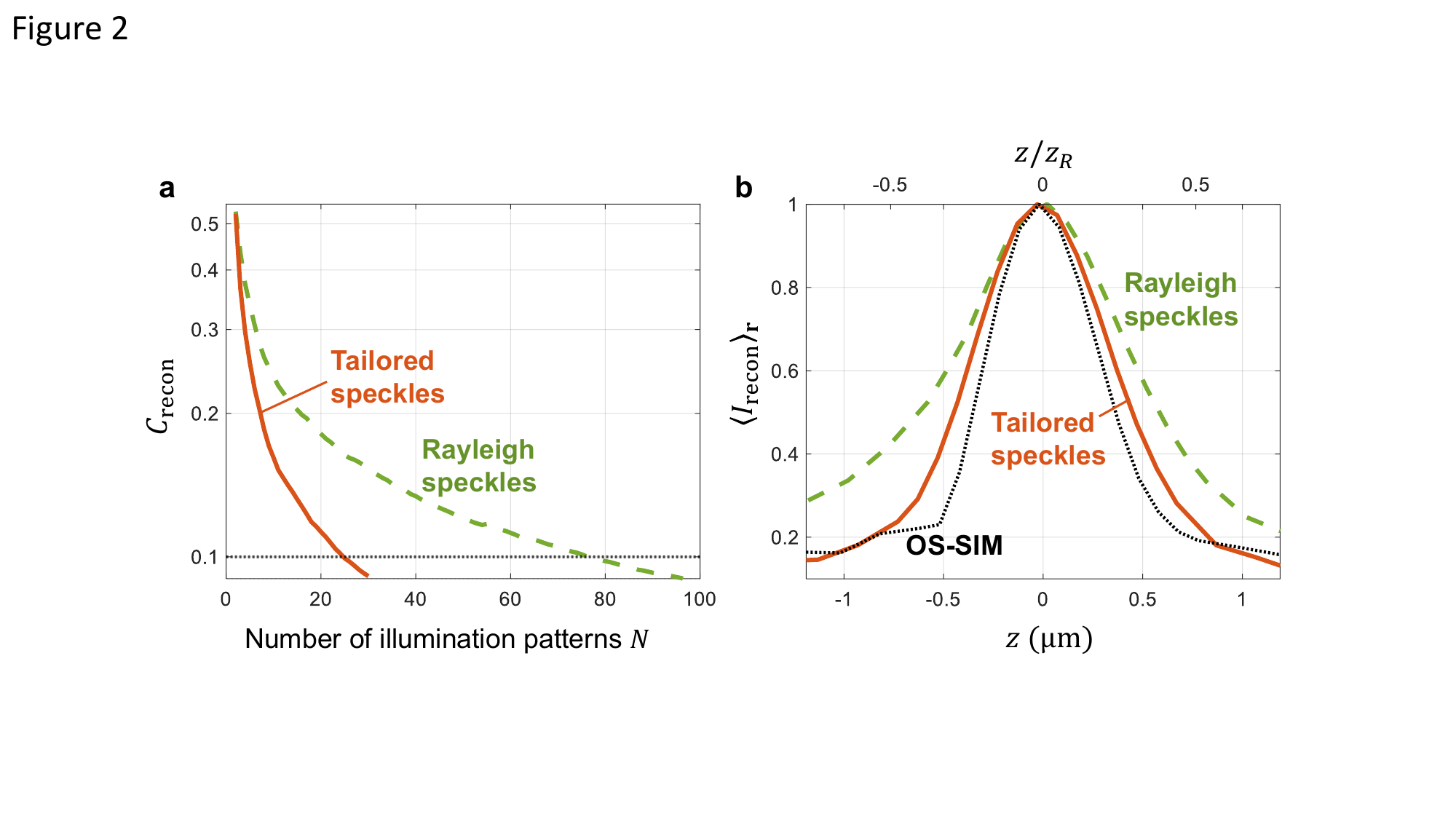} 
			\caption{\textbf{Image quality and axial sectioning of tailored speckles.}
			(a) Image reconstruction noise, quantified by the spatial contrast of reconstructed fluorescence intensity $I_{\mathrm{recon}}$ of a uniform thin fluorescent film, decreases with the number of uncorrelated illumination speckle patterns $N$. The black dashed line marks the spatial contrast of 0.1, which is reached with $N = 25$ tailored speckle patterns and $N = 77$ Rayleigh speckle patterns. (b) Reconstructed fluorescence signal as a function of axial position of the thin fluorescent film obtained by DSI with Rayleigh speckles (green dashed), tailored speckles (orange solid) and OS-SIM with periodic fringes (black dotted). Tailored speckle illumination improves axial sectioning relative to Rayleigh speckles. }\label{fig_sectioning_and_RMScontrast}
		\end{figure*}
		
		\subsection{Enhancing axial sectioning}
		
		Next we customize out-of-focus speckle statistics to increase axial sectioning strength. As mentioned earlier, the DSI sectioning capability relies on the reduction of the observed fluorescence intensity fluctuations when out of focus. In conventional DSI, this reduction arises solely from the blurring properties of the detection PSF. To improve axial sectioning, we can further reduce the fluctuations by lowering the intensity contrast of illumination itself when out of focus. As seen in Figure~\ref{fig_schematic}a, the out-of-focus planes in B that are closer to focus S have a smaller number of speckle grains within the green shaded region and thus higher fluctuation of fluorescence intensity when illuminating with uncorrelated speckle patterns. Our aim is to lower the intensity contrast of illumination speckles in these planes but not in the in-focus plane. This can be realized by axial variation of speckle intensity contrast, so that the in-focus contrast is higher than that away from focus. We further require that the in-focus speckle has a binary intensity PDF to minimize image reconstruction noise as described in the previous subsection. 
		
		Engineering a single speckle field to exhibit distinct statistics at different axial planes is challenging, because the field distribution at a single plane determines fields at all other planes via the free-space propagation relation. However, in our case we tailor only the statistics of the intensity patterns at different planes and not the patterns themselves, providing flexibility in our field engineering approach. Recently we developed a method of tailoring 3D speckle intensity statistics by optimizing the field amplitude and phase distributions \cite{han2023tailoring}. The algorithm is modified to include a bound for spatial frequencies of the fields. It is described briefly in Methods, and more details are given in Sec.~\ref{sec:supp_field_optimizations} of the Supplementary Materials. 
		
		We obtain 3D speckle fields that satisfy a low-contrast intensity PDF in four planes at $z = \pm 0.7 z_R, \pm 1.4 z_R$ and a binary PDF in focus $z=0$. Here, $z_R$ denotes the full width at half maximum (FWHM) of axial intensity decorrelation of Rayleigh speckles (Fig.~\ref{fig_schematic}b). It also sets the axial propagation distance over which a binary speckle pattern would return to Rayleigh statistics. To prevent such evolution and instead reduce the speckle contrast in nearby planes, we impose low speckle contrast at $z = \pm 0.7 z_R$. To suppress speckle contrast increase immediately beyond these planes, we further require low speckle contrast in two planes $z = \pm 1.4 z_R$. An example of tailored speckles is shown in Fig.~\ref{fig_schematic}c. In focus is a binary speckle pattern with intensity contrast close to 1, and out of focus speckle contrast is about 0.5. For comparison, Rayleigh speckles have axially evolving intensity patterns but the intensity contrast remains 1 (Fig.~\ref{fig_schematic}d).

		Once the 3D speckle fields with tailored intensity statistics are found numerically, they will be generated by a spatial light modulator (SLM) placed at the conjugate plane of $z=0$ (see experimental setup in Methods). To control both amplitude and phase of the fields at $z=0$, we write a grating on the phase-only SLM and filter the first-order diffraction \cite{bolduc2013exact, goorden2014superpixel}. A gradient-based optimization scheme is implemented to find the SLM phase pattern that produces the target field distribution at $z=0$ (details in Sec. \ref{sec:supp_SLM_optimizations} in Supplementary Materials, and codes available on \href{https://github.com/SeungYunH/tailored-speckle-imaging}{Github}). After loading the optimized phase pattern to the SLM, we measure the intensity distribution at $z=0$ and obtain more than 94\% correlation with the target pattern.

		\begin{figure*}[b]
			\centering
			\includegraphics[width=0.95\textwidth,
			trim=50pt 115pt 70pt 112pt,clip]{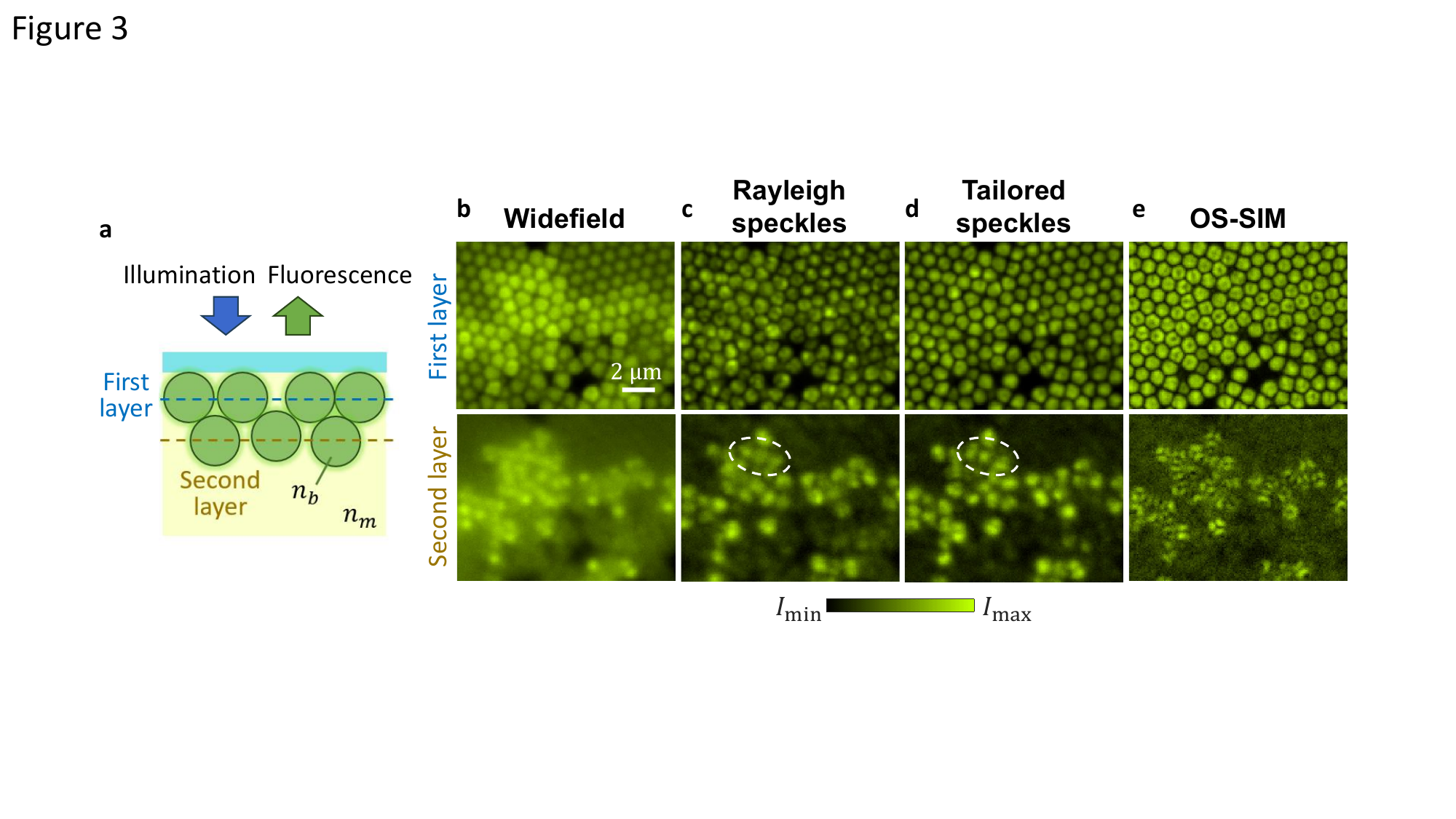} 
			\caption{\textbf{Tolerance of tailored speckles to scattering.}
			(a) Schematic of imaging fluorescent beads in epi configuration. Beads of 1 \textmu m diameter are stacked in two layers on a coverslip and immersed in water. Their index mismatch ($\Delta n = 0.26$) induces light scattering which distorts illumination patterns especially onto the second layer. (b–e) Reconstructed images of first and second layers when illuminated by  (b) widefield uniform pattern, (c) Rayleigh speckles, (d) tailored speckles, and (e) periodic fringes of OS-SIM. For the second layer, tailored speckles provide the best imaging quality with clear shapes of individual beads and strong axial sectioning. White scale bar of 2 \textmu m applies to all images.}\label{fig_beads}
		\end{figure*}
		\subsection{Experimental validation}		
		
		To verify the performance enhancement by tailored speckles experimentally, we prepare a uniform thin fluorescent film of Rhodamine 6G (see sample preparation in Methods). The fluorescent film is placed in focus, and illuminated at 488 nm with speckles of either Rayleigh or tailored intensity statistics. After recording the individual fluorescence images for $N$ different illuminations, we reconstruct the image $I_\mathrm{recon}(\mathbf{r})$ with Eq.~\ref{Eq:DSIrecon}. Note that the contributions of shot noise and camera noise are subtracted in  $I_\mathrm{recon}$ throughout the paper~\cite{ventalon2006dynamic}(see image reconstruction in Methods). Since the fluorescence is uniform, $I_\mathrm{recon}$ should be constant across the field of view (FoV), but finite sampling leads to residual speckle noise. We quantify the reconstruction noise with spatial contrast of $I_\mathrm{recon}$, $C_{\mathrm{recon}} \equiv {\sigma_{\mathbf r}[I_\mathrm{recon}(\mathbf r)]} / {\langle I_\mathrm{recon}(\mathbf r)\rangle_{\mathbf r}}$.
		
		Figure~\ref{fig_sectioning_and_RMScontrast}a shows a comparison of $C_{\mathrm{recon}}$ for Rayleigh and tailored speckles as a function of $N$. In both cases, $C_{\mathrm{recon}}$ decays as $N$ increases, but the decay rate is notably different. The tailored speckles yield much faster decay, which is attributed to the binary speckle statistics. To reach $C_{\mathrm{recon}} = 0.1$ (marked by dotted black line in Fig.~\ref{fig_sectioning_and_RMScontrast}a), we need $N = 25$ frames for tailored speckles and $77$ frames for Rayleigh speckles. The threefold reduction in the number of frames makes the imaging speed three times faster for the same reconstruction noise level.
		
		We also evaluate the axial sectioning ability of the tailored speckles by scanning the fluorescent film axially. Figure \ref{fig_sectioning_and_RMScontrast}b shows the FoV-averaged $\langle I_\mathrm{recon} \rangle_{\mathbf{r}}$ as a function of axial position $z$. Away from the focus $z=0$, $\langle I_\mathrm{recon} \rangle_{\mathbf{r}}$ decreases continuously. The FWHM of $\langle I_\mathrm{recon} \rangle_{\mathbf{r}}(z)$ is \qty{0.90}{\um} for tailored speckles and \qty{1.21}{\um} for Rayleigh speckles. The enhanced sectioning results from the lower speckle contrast away from focus. We further compare the optical sectioning with OS-SIM. For the strongest sectioning, the frequency of sinusoidal fringes for OS-SIM is set to half of the maximum spatial frequency of the detection optics \cite{li2025enhancing}. Figure \ref{fig_sectioning_and_RMScontrast}b shows the OS-SIM offers slightly better sectioning ability (black dotted curve), FWHM = \qty{0.72}{\um}, but it is more susceptible to light scattering than tailored speckles, as will be shown in the next section.
		

		\begin{figure*}[b]
			\centering
			\includegraphics[width=0.95\textwidth,
			trim=90pt 45pt 35pt 42pt,clip]{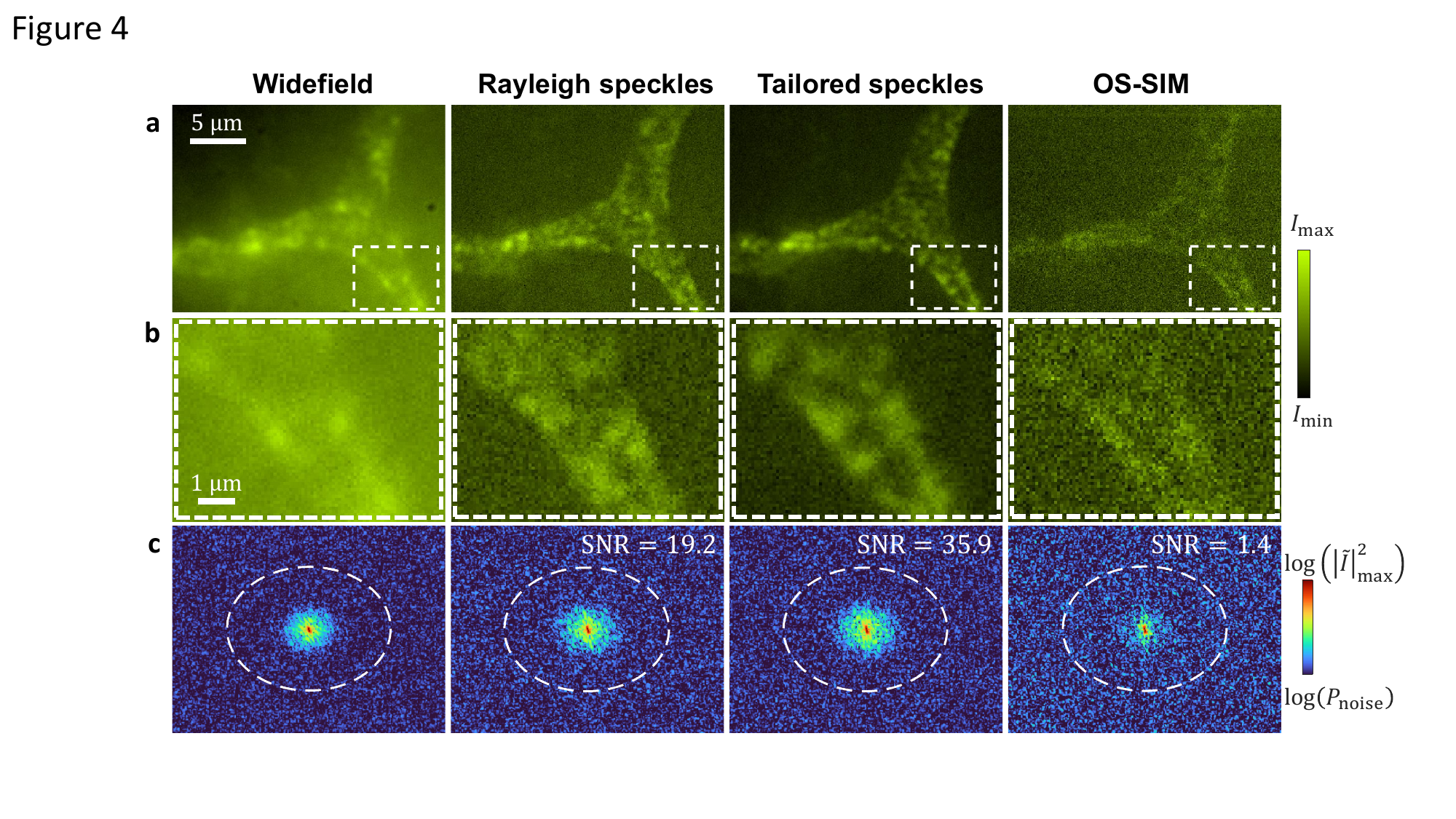} 
			\caption{\textbf{Mouse brain vascular imaging.}
			(a) Fluorescence images of the same field of view (18 \textmu m $\times$ 24 \textmu m) acquired with widefield imaging (first column), DSI with Rayleigh speckles (second column), tailored speckles (third column), and OS-SIM (fourth column). 
			The SNR values are 19.2 (Rayleigh speckles), 35.9 (tailored speckles), and 1.4 (OS-SIM). 
			(b) Magnified views of the dashed squares in (a), highlighting the region of strong background. DSI with tailored speckle provides the clearest image of the fine structure and lowest background. (c) Power spectra of the reconstructed images in (a) plotted in logarithmic scale. The dashed circle marks the OTF support defined by our detection optics. In DSI with tailored speckles, there are more high spatial frequency components above the noise level than in other schemes. The lower limit of the colorbar is set to the noise level.
			}\label{fig_vasculature}
		\end{figure*}

		\section{DSI with tailored speckles}
		
		\subsection{Scattering fluorescent beads}
		
		OS-SIM relies on well-defined fringe patterns that can be distorted by sample-induced aberration and scattering \cite{demmerle2017strategic}, thereby lowering the quality of reconstructed images \cite{mo2021structured}. In contrast, tailored speckle illumination is less dependent on specific intensity patterns and can therefore be more tolerant to distortions.
		
		To test the tolerance to sample-induced aberration and scattering, we image stacked layers of fluorescent beads (1 \textmu m diameter) on a coverslip. These polystyrene beads (refractive index $n_b=1.59$) are immersed in water ($n_m=1.33$), resulting in a notable index mismatch at the bead--water interfaces. This mismatch introduces appreciable light scattering and aberration. In the epi-geometry of our experiment, Fig.~\ref{fig_beads}a shows illumination through the coverslip and fluorescence collected from the same side. We refer to the bead layer attached to the coverslip as the "first layer" and the beads stacked above it as the "second layer." Because beads in the first layer introduce light scattering that distorts the excitation pattern on the second layer, beads in the second layer are more difficult to image than the first. 
		
		The double-layer structure of fluorescent beads allows direct comparison of different illumination patterns.	
		In widefield imaging of the first layer, due to the lack of axial sectioning, double-layered regions show brighter fluorescence than single-layered regions in Fig.~\ref{fig_beads}b. Imaging the second layer by moving it to focus also reveals an elevated background level originating from fluorescence of the first layer.
		
		DSI reconstruction with Rayleigh speckles achieves optical sectioning, as seen in Fig.~\ref{fig_beads}c. In the reconstructed image of the first layer, regions that have the second layer no longer appear as bright as in the widefield image, indicating effective depth discrimination. However, some beads are brighter than others, and even within single beads fluorescence can be non-uniform, due to reconstruction noise with finite ($N = 30$) different speckle illuminations. 
		Nevertheless, Rayleigh speckle statistics do not change by light scattering and the second-layer reconstruction shows appreciable depth discrimination, suppressing background fluorescence from the first layer.	This highlights the key advantage of optical sectioning with random speckle illumination: intrinsic robustness to scattering and aberration.
		
		Tailored speckle illumination improves the image quality over that with Rayleigh speckles (Fig.~\ref{fig_beads}d). For the same number of illumination patterns $N = 30$, reconstruction noise is reduced. The first-layer image shows more uniform fluorescence across the FoV and within individual beads. In the second-layer reconstruction, the bead boundaries become clearer (white dashed region), and the background fluorescence from the first-layer beads is further suppressed. These results demonstrate that customized speckle statistics remain effective even in the presence of appreciable sample-induced light scattering.
		
		In contrast, the OS-SIM reconstruction in Fig.~\ref{fig_beads}e is strongly degraded by the sample scattering. In the first-layer image, although the fluorescence remains relatively uniform across the FoV, individual beads in double-layered regions exhibit reconstruction artifacts (dark spots at the center). The degradation of OS-SIM reconstruction becomes more severe when imaging the second layer. The bead shapes and boundaries are no longer reliably reconstructed. We attribute these artifacts to sample-induced light scattering, which distorts the excitation fringe patterns. 
		To confirm that this degradation arises from sample-induced scattering, we replace water with an index-matched fluid, which restores the image quality with OS-SIM (Sec.~\ref{sec:supp_Index_match} of the Supplementary Materials). We further implement incoherent OS-SIM to compare with coherent OS-SIM. Although incoherent OS-SIM has stronger axial sectioning, it exhibits even worse image reconstruction of double-layered beams than coherent OS-SIM due to reduction of fringe depth (Sec.~\ref{sec:supp_incSIM} of the Supplementary Materials).

		\subsection{Mouse brain vascular imaging}
		
		Finally we apply DSI with tailored speckles to imaging in an optically thick biological specimen. A fixed mouse brain slice is labeled for blood vessels using fluorescent lectin (see sample preparation in Methods). Figure~\ref{fig_vasculature} shows fluorescent images of blood vessels at a depth of 25~\textmu m using a 40$\times$, NA=0.95 air objective lens. The excitation laser is at a wavelength of 488 nm as in the previous experiments, and the fluorescence is centered at 519 nm. 
		
		The widefield image captures the overall vascular structure, but some regions have a strong background from other depths, making it difficult to discern fine structures in the magnified view as shown in Fig.~\ref{fig_vasculature}b. DSI with Rayleigh speckles provides an optically sectioned image that separates the blood vessels from the background much more clearly. Tailored speckle illumination further improves the image quality, yielding sharper boundaries. In contrast, OS-SIM yields much lower image quality due to sample-induced scattering and aberration. 
		
		We further perform a spatial Fourier transform of reconstructed images in Fig.~\ref{fig_vasculature}a, and show the power spectra in Fig.~\ref{fig_vasculature}c. The white dashed circle marks the range of spatial frequency of the optical transfer function (OTF). Compared to OS-SIM, DSI with tailored speckles reconstructs more signal at high spatial frequencies under the same experimental conditions. The spatial-frequency spectra further reveal that tailored speckle illumination preserves the spatial resolution in the reconstructed image, despite the illumination pattern itself having weaker high-spatial-frequency components than Rayleigh speckles (see Sec.~\ref{sec:supp_powerSpectrum} of the Supplementary Materials).
        
		
		For quantitative comparison, we estimate the signal-to-noise ratio (SNR) in Fig.~\ref{fig_vasculature}c. The frequency components outside the OTF support (white dashed circle) come from detection noise. Assuming the noise level $P_{\mathrm{noise}}$ is constant over the spatial frequency, we estimate it as $P_{\mathrm{noise}} = \langle P\rangle_{\mathrm{outside}}$, the mean outside the OTF support. The mean value inside the OTF support is then the sum of signal and noise, $\langle P\rangle_{\mathrm{inside}}=P_{\mathrm{signal}} + P_{\mathrm{noise}}$. Assuming the noise level inside the OTF support is identical to that outside, the signal is then estimated as $P_{\mathrm{signal}} = \langle P\rangle_{\mathrm{inside}}-P_{\mathrm{noise}}$, and we define the SNR as $P_{\mathrm{signal}}/P_{\mathrm{noise}}$. As written in Fig.~\ref{fig_vasculature}c, SNR = 35.9 for DSI with tailored speckles, approximately $2\times$ higher than that with Rayleigh speckles (SNR = 19.2), and OS-SIM has the lowest SNR of 1.4.  
		
		
		\section{Discussion and conclusion}
			
		In our current setup with a high-NA air objective, the original DSI based on Rayleigh speckles reconstructs images up to \qty{40}{\um} in the mouse brain sample. This depth limit is similar to the scattering mean free path of mouse brain cortex ($l_s = 38 \pm 2$ \textmu m at $\lambda$ = 515 nm) \cite{imperato2022single}. Although Rayleigh speckle statistics is not affected by strong light scattering, the detection PSF is degraded, limiting the imaging depth.  
		Near the imaging depth limit (\qty{40}{\um}), DSI with tailored speckles still provides higher SNR and stronger optical sectioning than that with Rayleigh speckles. This result suggests that the tailored speckle statistics is maintained at least partly near the imaging depth limit set by detection PSF degradation. This limit can be extended by reducing the refractive-index mismatch between the objective immersion medium and the sample, for example by using a water-immersion objective. Adaptive optics techniques may be adopted to further increase the imaging depth.
		
		In conclusion, we tailor speckle intensity statistics for non-scanning fluorescence imaging with optical sectioning. By simultaneously imposing binary speckles in focus and low speckle contrast out of focus, we reduce image reconstruction noise and strengthen optical sectioning. Our scheme remains effective under sample-induced light scattering and aberrations. Images of index-mismatched bead layers and mouse brain blood vessels reveal that customized speckles outperform Rayleigh speckles that are commonly used for DSI microscopy, and preserve image quality in scattering samples where OS-SIM suffers severe degradation.
		
		While we optimize the speckle intensity statistics for fluorescence imaging under linear excitation here, our method can be applied to other imaging modalities. For example, speckled illuminations have been used for multi-photon fluorescence imaging \cite{negash2018two, zhu2025two}, stimulated Raman scattering (SRS) \cite{hokr2017enabling}, and coherent anti-Stokes Raman scattering (CARS) microscopy \cite{heinrich2008coherent, fantuzzi2023wide}. Customizing speckle statistics for these applications will optimize the imaging performance, e.g., minimizing speckle noise, maximizing axial resolution. Furthermore, super-resolution microscopy that uses random illumination may also benefit from tailored speckles to improve image quality \cite{mudry2012structured, mangeat2021super, yoo2025near}.

		\section*{Methods}
		
		\subsection*{Sample preparation}
		\vspace{-7pt}
		\noindent Thin film of fluorescent dye. Rhodamine 6G is dissolved in ethanol and spin-coated onto a coverslip. The layer thickness is estimated to be less than  \qty{1}{\nm}, which is much smaller than the free-space axial resolution of our imaging system.
		
		\vspace{10pt}
		\noindent Bilayer of fluorescent beads.
		Fluorescent microspheres with Nile Red are deposited onto the No. 1.5H coverslip by drying a droplet of bead suspension. The coverslip with stacked beads is then mounted onto the sample stage of our imaging setup. Immediately before the imaging experiment, the beads are immersed in a droplet of deionized water or index-matching fluid.
		
		\vspace{10pt}
		\noindent Mouse brain sample.  
		Male C57BL/6 mice (12 months old) are anesthetized and live transcardial perfusion is performed using phosphate-buffered saline (PBS), followed by paraformaldehyde (PFA) to preserve tissue integrity. The brains are then extracted and post-fixed in PFA. After washing the fixed tissue, coronal sections (\qty{150}{\um} thick) are prepared using a vibratome. For vascular labeling, brain sections are incubated with tomato lectin conjugated with DyLight 488 in PBST. The sections are subsequently washed with PBS and mounted on a glass slide with PBS using a spacer and covered with a No. 1.5H coverslip.
		
		More details on sample preparations can be found in Supplementary Sec.~\ref{sec:supp_sample}.
		
		\subsection*{Optical setup}
		\vspace{-7pt}
		We built a custom fluorescence microscope in which a phase-only SLM (X10468-01, LCOS-SLM, Hamamatsu) shapes the \qty{488}{\nm} excitation wavefront to generate target illumination patterns on the sample. A blazed grating is added to the phase modulation pattern on the SLM to generate multiple diffraction orders, and a Fourier-plane iris selects the first order, enabling complex-field modulation at the sample with a phase-only SLM. The filtered beam is imaged onto the sample by an air objective (40$\times$, NA = 0.95, Plan apo $\lambda$D, Nikon).
		
		Fluorescence is collected in epi-geometry through the same objective and recorded on a camera (DU940P-BV, Newton CCD, Andor). An optical beam shutter (SH05/M, Thorlabs) is placed in the excitation beam path and synchronized with the camera. The illumination power in the sample plane ($z=0$) varies from \qty{116}{\uW} for a plane wave to \qty{58}{\uW} for the bespoke speckles. The camera exposure times per frame are \qty{0.7}{\s} for thin dye film, and \qty{0.05}{\s} for fluorescent beads and mouse brain blood vessels. A schematic of our optical setup and additional details are provided in Supplementary Sec.~\ref{sec:supp_setup}.
		
		\vspace{-10pt}
		
		\subsection*{Tailored speckle generation}
		\vspace{-7pt}		 
		
		Unlike our previous method \cite{han2023tailoring}, the phase-only SLM is placed in a conjugate plane of the sample plane $z=0$. We numerically optimize the speckle field at $z=0$ to obtain the target intensity PDFs in five axial planes at $z = 0, \pm 0.7 z_R, \pm 1.4 z_R$. This optimization utilizes a Gerchberg-Saxton-like algorithm that bounds the spatial frequencies of fields by the numerical aperture (NA) of illumination optics. Details can be found in Sec. \ref{sec:supp_field_optimizations} in Supplementary Materials, and codes available in \href{https://github.com/SeungYunH/tailored-speckle-imaging}{Github}. 
		
		Next, we create the target speckle field at $z=0$ by optimizing the SLM phase pattern. With a forward model of field propagation from the SLM to the sample plane, including the first-order filtering, the SLM phase pattern is optimized by a gradient-descent-based algorithm to generate the target field at $z=0$. The SLM pixels are no longer grouped into macro-pixels, unlike in our previous work \cite{han2023tailoring}. More details are given in the Supplementary Sec. \ref{sec:supp_SLM_optimizations}, and codes available in \href{https://github.com/SeungYunH/tailored-speckle-imaging}{Github}. Our SLM with $600\times792$ pixels can generate custom fields in a sample area of \qtyproduct{37 x 49}{\um} that contains about $1.5\times10^{4}$ speckle grains. Experimentally, we reduce the FoV to \qtyproduct{18 x 24}{\um} that has $4\times10^{3}$ speckle grains. Full-FoV tailored speckle imaging ($1.5\times10^{4}$ speckle grains) gives similar performance to reduced FoV, as shown in Supplementary Sec.~\ref{sec:supp_FullFOV}
.		
		\subsection*{Image reconstruction}
		\vspace{-7pt}
		
		For a given set of $N$ illumination patterns (speckles for DSI and sinusoidal fringes for OS-SIM), the corresponding fluorescence images captured by camera $I(\mathbf{r})$ are used to calculate the intensity variance over frames:
		\begin{equation}
			V(\mathbf{r})= \left\langle I(\mathbf{r})^2 \right\rangle_{N} -  \left\langle I(\mathbf{r}) \right\rangle^2_{N},
		\end{equation}
		where $\left\langle \cdot \right\rangle_{N}$ means the mean over $N$ frames. To account for shot noise and camera noise, the camera gain $G$ (counts/e$^{-}$) and noise variance $\sigma_{\mathrm{read}}^{2}$ (counts$^{2}$) are calibrated \cite{ventalon2006dynamic}. The reconstructed signal is:
		\begin{equation}
			I_{\mathrm{recon}}(\mathbf{r})=\sqrt{V(\mathbf{r})-G \left\langle I(\mathbf{r}) \right\rangle_{N}-\sigma_{\mathrm{read}}^{2}}.
		\end{equation}
		\noindent Any negative values after subtraction are clipped to zero before taking the square root.
		
		DSI with Rayleigh speckles and tailored speckles both use $N=30$ uncorrelated illumination patterns in Figs.~\ref{fig_beads} and \ref{fig_vasculature}. OS-SIM uses three lateral shifts ($N=3$) of one fringe orientation to reconstruct an image. This procedure is repeated for 10 orientations with angles sampled uniformly in $[0,\pi]$. Reconstructed images from 10 orientations are then averaged into a single OS-SIM image with a total of 30 frames.
		
		Finally, we multiply all reconstructed images by an intensity envelope to correct for nonuniform illumination across the field of view, caused by position-dependent first-order diffraction efficiency.
		
		\bmsection*{Acknowledgments}
		We thank Nathan Vigne, Rohin McIntosh, and Mert Ercan for insightful discussions.
		
		\bmsection*{Funding Statement}
        This research was supported partly by the National Institutes of Health (NIH) under Grant No. R01GM160992 and by the Chan Zuckerberg Initiative DAF (an advised fund of the Silicon Valley Community Foundation) under grant 2021-234544; and the Bio\&Medical Technology Development Program of the National Research Foundation (NRF), funded by the Korean government (MSIT), under grant RS-2023-00264980.
		
		\bmsection*{Competing Interests}
		The authors have no competing interests.
		
		\bmsection*{Data Availability Statement}
		The data supporting the findings of this study are available from the corresponding author upon reasonable request. Optimization codes are available: \href{https://github.com/SeungYunH/tailored-speckle-imaging}{https://github.com/SeungYunH/tailored-speckle-imaging}

		\bmsection*{Ethical Standards}
		The research meets all ethical guidelines, including adherence to the legal requirements of the study country.
		
		\bmsection*{Author Contributions}
		S.H. built and calibrated the optical setup, prepared samples, developed the field-optimization algorithm, derived the binary speckle equations together with K.W., acquired and analyzed the data, and wrote the manuscript. K.L. contributed to the setup construction, developed the SLM pattern-optimization algorithm, and provided scientific discussion and manuscript revision. Y.S.K. prepared the mouse brain samples and wrote the relevant Methods section. C.L. performed the literature search and reviewed the manuscript. N.B. contributed to the conceptualization and scientific discussion. K.W. contributed to the derivation of the binary intensity statistics. T.K. supervised the preparation of the mouse brain samples. J.M. provided critical advice on experiments and datasets and reviewed the manuscript. H.C. conceived the idea, supervised the project, and reviewed the manuscript. All authors discussed the results and approved the final manuscript.

		
		\bibliography{main}
		
	\end{document}


	\onecolumn
	\begin{center}
		{\Large\bfseries Supporting Information for:\\
			Tailored Speckle Illumination Microscopy with Enhanced Sectioning and Image Quality\par}
		\vspace{0.6em}
		
		SeungYun Han$^{1}$, KyeoReh Lee$^{1}$, Young Seo Kim$^{2}$, Chuan Li$^{3,4}$,\\
		Nicholas Bender$^{5}$, Kabish Wisal$^{6}$, Taeyun Ku$^{2}$, Jerome Mertz$^{4,7}$, Hui Cao$^{1}$\par
		\vspace{0.6em}
		
		{
			$^{1}$Department of Applied Physics, Yale University, New Haven, CT 06510, USA\\
			$^{2}$Graduate School of Medical Science and Engineering, KAIST, Daejeon 34141, Republic of Korea\\
			$^{3}$Department of Electrical \& Computer Engineering, Boston University, Boston, MA 02215, USA\\
			$^{4}$Photonics Center, Boston University, Boston, MA 02215, USA\\
			$^{5}$School of Applied \& Engineering Physics, Cornell University, Ithaca, NY 14853, USA\\
			$^{6}$Department of Physics, Yale University, New Haven, CT 06510, USA\\
			$^{7}$Department of Biomedical Engineering, Boston University, Boston, MA 02215, USA\\
			Corresponding author: Hui Cao (\texttt{hui.cao@yale.edu})
		}
	\end{center}
	
	\vspace{1em}
	
	\renewcommand{\thefigure}{S\arabic{figure}}
	\renewcommand{\thetable}{S\arabic{table}}
	\renewcommand{\theequation}{S\arabic{equation}}
	\setcounter{figure}{0}
	\setcounter{table}{0}
	\setcounter{equation}{0}
	
	\renewcommand{\thesection}{S\arabic{section}}
	\renewcommand{\thesubsection}{S\arabic{section}.\arabic{subsection}}
	\renewcommand{\thesubsubsection}{S\arabic{section}.\arabic{subsection}.\arabic{subsubsection}}
	
	\setcounter{section}{0}

	\vspace{-40pt}

	\section{Sample preparation} \label{sec:supp_sample}
	
	\subsection{Thin film of fluorescent dye}
	
	Rhodamine 6G (Exciton) was dissolved in ethanol and diluted to \qty{2.5}{\ug/mL}. The solution was spin-coated onto a coverslip (22$\times$22~mm, No. 1.5H, Thorlabs) using a spin coater (Laurell Technologies). For spin coating, \qty{100}{\uL} of the solution was dispensed onto the coverslip and spun at 4000~rpm for 1~min. Roughly, the total mass of \qty{0.25}{\ug} (with density \qty{1.26}{\g/cm^3}) is spread over \qty{4.84}{\cm^2}. The layer thickness of fluorescent dye is less than \qty{1}{\nm} on average, which is much smaller than the free-space axial resolution of the imaging system (\qty{0.71}{\um}).

	\subsection{Bilayer of fluorescent beads}
	
	Fluorescent microspheres (FluoSpheres, Nile Red, Invitrogen; nominal diameter 1.0~µm) are deposited onto the No. 1.5H coverslip by drying a droplet of bead suspension \qty{5}{\uL} on the glass surface for 10 minutes. The weakly adhered beads are then removed by rinsing the coverslip several times with deionized (DI) water. After drying, the coverslip with remaining beads is mounted on the sample stage of our imaging setup. Immediately before the imaging experiment, the beads are immersed in DI water or index-matching fluid by placing a droplet on the coverslip. These procedures minimize bead detachment with time and thereby prevent floating beads in solution which would generate dynamic background fluorescence.

	\subsection{Mouse brain sample}
	  
	Animal care and experimental procedures are performed under approval from the Institutional Animal Care and Use Committee (IACUC) of the Korea Advanced Institute of Science and Technology (KAIST). All experimental procedures in this study are conducted in accordance with the IACUC-approved guidelines. Male C57BL/6 mice (12 months old) are housed with free access to food and water. For tissue preparation, mice are anesthetized via intraperitoneal injection of Avertin (2,2,2-tribromoethanol; T48402, Sigma-Aldrich) at a dosage of \qty{20}{\uL\per\gram} body weight. Live transcardial perfusion is performed using \qty{15}{\mL} of phosphate-buffered saline (PBS), followed by \qty{15}{\mL} of 4\% paraformaldehyde (PFA; 15714-S, Electron Microscopy Sciences) in PBS to preserve tissue integrity. Following perfusion, the brains are extracted and post-fixed in \qty{35}{\mL} of 4\% PFA at 4℃ overnight, followed by an additional 12 h at room temperature. The fixed tissue is then washed three times with \qty{35}{\mL} of PBS at 12 h intervals at 4℃. Coronal sections (\qty{150}{\um} thick) are prepared using a vibratome (VT1200S, Leica). For vascular labeling, brain sections are incubated with \qty{1}{\uL} of tomato lectin conjugated with DyLight 488 (L32470, Thermo Fisher Scientific) in \qty{300}{\uL} of PBST (0.1\% Triton X-100 in PBS containing 0.02\% sodium azide) at 37℃ overnight. The sections are subsequently washed three times with PBS at intervals of 30 min. Finally, the tissue is mounted on a glass slide with PBS using a spacer and covered with a No. 1.5H coverslip (Matsunami).

	\section{Optical setup} \label{sec:supp_setup}
	
\begin{figure*}[htpb]
		\centering
		\includegraphics[width=0.7\textwidth,
		trim=155pt 120pt 230pt 162pt,clip]{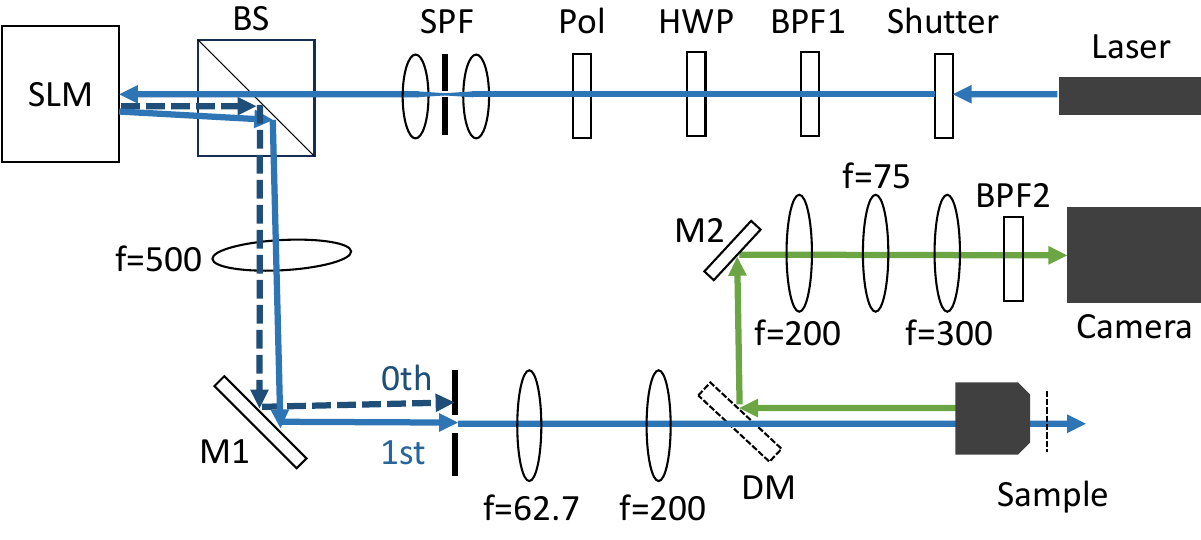} 
		\caption{\textbf{Schematic of optical setup.}
			A monochromatic laser beam at $\lambda$ = 488 nm (blue) is spectrally and spatially filtered then linearly polarized before projecting onto a phase-only SLM. A $2f_1-2f_2$ relay with filtering of the first-order diffraction by an iris in the Fourier plane realize complex-field modulation at the sample plane. Fluorescence from the sample (green) is collected in epi-geometry through the same objective, separated by a dichroic mirror (DM), spectrally filtered by a band-pass filter (BFP2), and imaged onto the CCD camera with another  $2f_1-2f_2$ relay for magnification. BS, beamsplitter; SPF, spatial filter; Pol, polarizer; HWP, half-wave plate; BPF1/BFP2, band-pass filters; M1/M2, mirrors. Focal lengths are indicated in the schematic (unit = mm).}\label{fig_supp_setup}
	\end{figure*}

	Fig.~\ref{fig_supp_setup} is a schematic of our optical setup for tailored speckle illumination microscopy. A 488~nm laser beam (50~mW, Sapphire series, Coherent) first passes through a mechanical shutter and a band-pass filter centered at 488~nm with a 3~nm passband (BPF1). Its polarization is adjusted with a half-wave plate (HWP) and a linear polarizer (Pol). The beam is spatially cleaned by a spatial filter (SPF) and expanded to fill the active area (\qty{15.84} \times \qty{12.00}{\mm^2}) of a phase-only spatial light modulator ($792 \times 600$ pixels, \qty{20}{\um} pixel pitch).
	
	The light reflected from the SLM is directed by a beamsplitter (BS) into the illumination arm. The SLM is placed in the front focal plane of a lens with focal length $f=500$~mm. A blazed grating (period $\sim$ 5 pixels) is written on the SLM to produce multiple diffraction orders in the back focal plane of the lens (Fourier plane). In this plane, an iris blocks all orders except the first order. The diameter of the iris sets the beam size so that it matches the objective lens back aperture downstream. The selection of first-order diffraction enables complex-field modulation at the sample plane ($z=0$) using a phase-only SLM. The filtered beam is then relayed by a self-imaging system ($2 f_1 - 2 f_2$, $f_1=62.7$~mm and $f_2=200$~mm) and coupled to an air objective (40$\times$, NA~=~0.95, Plan apo $\lambda$D, Nikon). 
	
	The objective is oriented horizontally and the sample is mounted vertically. The objective images the sample through the coverslip, and the correction ring is adjusted for the thickness of the coverslip. The sample is brought into focus by first using a motorized translation stage (\qty{12}{\mm} travel range, Thorlabs) for coarse positioning, followed by fine focusing with a piezo stage (P-753, Physik Instrumente). We translate the sample laterally using a manual translation stage.
	
	Fluorescence from the sample is collected in epi-geometry and directed to the detection arm by a 505-nm shortpass dichroic mirror (DC, DMSP505R, Thorlabs). The fluorescence is reflected by a mirror (M2), imaged by a lens ($f=200$~mm), relayed by a magnifying self-imaging system ($2 f_1 - 2 f_2$, $f_1=75$~mm and $f_2=300$~mm), spectrally filtered by a band-pass filter centered at 535~nm with a 70~nm bandwidth (BPF2), and recorded by the camera (512 × 2048 pixels, 485 × 633 pixels used for FoV, \qty{13.5}{\um} pixel pitch, DU940P-BV, Newton CCD, Andor). The exposure times per frame are 0.7~s for the thin fluorescent film and 0.05~s for fluorescent beads and mouse brain blood vessels.
	
	\section{Binary speckles}
	\label{sec:supp_binary_derivation}
	
	Here, we derive Eq. (5) in the main text. As mentioned in the main text, we assume that (i) each illumination $M_n(\mathbf r)$ is independently sampled from the same intensity distribution, and that each frame contains sufficiently many speckle grains such that (ii) spatial averages of statistical quantities are identical for all $n$. The illumination intensity is standardized as Eq. (2) in the main text,
	\begin{equation}
		J_n(\mathbf r)=\frac{
			M_n(\mathbf r)-\left\langle M_n(\mathbf 
			r)\right\rangle_{\mathbf r}
		}
		{
			\sigma_{\mathbf{r}}\!\left( M_n(\mathbf r)\right)
		},
	\end{equation}
	where $\langle\cdot\rangle_{\mathbf r}$ and $\sigma_{\mathbf{r}}(\cdot)$ denotes the average and standard deviation over $\mathbf r$, respectively. Consequently, $ \langle J_n (\mathbf r)\rangle_\mathbf r = 0$ and $ \langle J_n ^2(\mathbf r)\rangle_\mathbf r = 1$.
	
	The variance across $N$ illuminations, $X(\mathbf r)$, is defined as $X(\mathbf r) \equiv \mathrm{Var}_{N}\!\big(J_n(\mathbf r)\big) = \langle J_n ^{2}(\mathbf r)\rangle_N - \langle J_n (\mathbf r)\rangle_N^{2}$. We write $X=A-B$ with
	\begin{equation}
		A=\frac{1}{N}\sum_{n=1}^N J_n ^{2}(\mathbf r),\qquad B=\left(\frac{1}{N}\sum_{n=1}^NJ_n(\mathbf r)\right)^2.
	\end{equation}
	
	\noindent First, we calculate $\langle X\rangle_{\mathbf r} = \langle A\rangle_{\mathbf r} - \langle B\rangle_{\mathbf r}$. With the assumption (ii) above, each term is calculated as follows:
	\begin{align}
		\langle A\rangle_{\mathbf r} &= \frac{1}{N}\sum_{n=1}^N \langle J_n ^{2}(\mathbf r) \rangle_{\mathbf r} = \langle J_n ^{2}(\mathbf r) \rangle_{\mathbf r} \\
		\langle B\rangle_{\mathbf r} &= \left\langle\frac{1}{N^2}\left(\sum_{n=1}^NJ_n(\mathbf r) \right)^2 \right\rangle_{\mathbf r} \\
		&= \frac{1}{N^2}\left\langle \sum_{n=1}^N J_n^2(\mathbf r)
		+ \sum_{n=1}^{N}\sum_{\substack{m=1\\ m\neq n}}^{N} J_n(\mathbf r)\, J_m(\mathbf r)\right\rangle_{\raisebox{9pt}{$\scriptstyle \mathbf r$}} \\
		&=\frac{1}{N}\langle J_n ^{2}(\mathbf r) \rangle_{\mathbf r} + \frac{N-1}{N}\langle J_n (\mathbf r) \rangle_{\mathbf r}^2 \\
		&= \frac{1}{N},    
	\end{align}
	where we used the condition $\langle J_n(\mathbf r)\rangle_{\mathbf r}~=~0$ and $ \langle J_n ^2(\mathbf r)\rangle_\mathbf r = 1$. Consequently, we get $\langle X\rangle_{\mathbf r} = \frac{N-1}{N}$.
	
	Next, we evaluate $\langle X^{2}\rangle_{\mathbf r}=\langle A^{2}\rangle_{\mathbf r}-2\langle AB\rangle_{\mathbf r}+\langle B^{2}\rangle_{\mathbf r}$. 
	For brevity, we use the shorthands
	\[
	\sum_{n\neq m}\equiv \sum_{n=1}^{N}\sum_{\substack{m=1\\ m\neq n}}^{N},
	\qquad
	\sum_{n\neq m\neq l}\equiv \sum_{n=1}^{N}\sum_{\substack{m=1\\ m\neq n}}^{N}\sum_{\substack{l=1\\ l\neq n,m}}^{N}, 
	\qquad
	\sum_{n\neq m\neq l\neq k}\equiv \sum_{n=1}^{N}\sum_{\substack{m=1\\ m\neq n}}^{N}\sum_{\substack{l=1\\ l\neq n,m}}^{N}\sum_{\substack{k=1\\ k\neq n,m,l}}^{N}.
	\]
	Then each term is calculated as follows:
	\begin{align}
		\langle A^2\rangle_{\mathbf r} &= \left\langle\frac{1}{N^2}\left(\sum_{n=1}^NJ_n^2(\mathbf r) \right)^2 \right\rangle_{\mathbf r} \\
		&= \frac{1}{N^2}\left\langle \sum_{n=1}^N J_n^4(\mathbf r)
		+ \sum_{n\neq m} J_n^2(\mathbf r)\, J_m^2(\mathbf r)\right\rangle_{\mathbf r} \\
		&=\frac{1}{N}\langle J_n ^{4}(\mathbf r) \rangle_{\mathbf r} + \frac{N-1}{N}\langle J_n^2 (\mathbf r) \rangle_{\mathbf r}^2 \\
		\langle AB\rangle_{\mathbf r}
		&= \left\langle \frac{1}{N^3}
		\left(\sum_{n=1}^N J_n^2(\mathbf r)\right)
		\left(\sum_{m=1}^N J_m(\mathbf r)\right)^2
		\right\rangle_{\mathbf r} \\
		&= \frac{1}{N^3}\left\langle
		\left(\sum_{n=1}^N J_n^2(\mathbf r)\right)
		\left(\sum_{m=1}^N J_m^2(\mathbf r)
		+ \sum_{m\neq l} J_m(\mathbf r)J_l(\mathbf r)\right)
		\right\rangle_{\mathbf r} \\
		&= \frac{1}{N^3}\left\langle
		\sum_{n=1}^N\sum_{m=1}^N J_n^2(\mathbf r)\, J_m^2(\mathbf r)
		+ \sum_{n=1}^N \sum_{m\neq l}
		J_n^2(\mathbf r)\, J_m(\mathbf r)\, J_l(\mathbf r)
		\right\rangle_{\mathbf r} \\
		&= \frac{1}{N^3}\left\langle
		\sum_{n=1}^N J_n^4(\mathbf r)
		+ \sum_{n \neq m } J_n^2(\mathbf r)\, J_m^2(\mathbf r)
		+ 2\sum_{n \neq m } J_n^3(\mathbf r) J_m(\mathbf r) +\sum_{n \neq m \neq l} J_n^2(\mathbf r) J_m(\mathbf r) J_l(\mathbf r)
		\right\rangle_{\mathbf r} \\
		&= \frac{1}{N^2}\langle J_n^{4}(\mathbf r)\rangle_{\mathbf r}
		+ \frac{N-1}{N^2}\langle J_n^{2}(\mathbf r)\rangle_{\mathbf r}^{2}. \\
		\langle B^2\rangle_{\mathbf r}
		&= \left\langle \frac{1}{N^4}
		\left(\sum_{n=1}^N J_n(\mathbf r)\right)^4
		\right\rangle_{\mathbf r} \\
		&= \frac{1}{N^4}\left\langle \begin{aligned}
			&\sum_{n=1}^N J_n^4(\mathbf r)
			+ 4\sum_{n \neq m} J_n^3(\mathbf r) J_m(\mathbf r)
			+ 3\sum_{n \neq m} J_n^2(\mathbf r) J_m^2(\mathbf r) + 6\sum_{n \neq m \neq l} J_n^2(\mathbf r) J_m(\mathbf r) J_l(\mathbf r)  + \sum_{n \neq m \neq l \neq k} J_n(\mathbf r) J_m(\mathbf r) J_l(\mathbf r) J_k(\mathbf r)
		\end{aligned} \!\!\!\!\!\right\rangle_{\mathbf r} \\
		&= \frac{1}{N^3}\langle J_n^{4}(\mathbf r)\rangle_{\mathbf r}
		+ \frac{3(N-1)}{N^3}\langle J_n^{2}(\mathbf r)\rangle_{\mathbf r}^{2}.
	\end{align}
	The cross term proportional to $\langle J_n^{2}(\mathbf r)\rangle_{\mathbf r}^{2}$ in $\langle B^2\rangle_{\mathbf r}$ can also be understood combinatorially; in the fourth-order expansion, we choose two factors to carry index $m$ and the remaining two to carry index $n$, giving $\frac{1}{N^4}\binom{4}{2}\binom{N}{2}\langle J_n^{2}(\mathbf r)\rangle_{\mathbf r}^{2}$. Combining the results, we get
	\begin{align}
		\langle X^2\rangle_{\mathbf r} &= \langle A^2\rangle_{\mathbf r} - 2\langle AB\rangle_{\mathbf r} + \langle B^2\rangle_{\mathbf r} \\
		&= \left[\frac{1}{N}\langle J_n ^{4}(\mathbf r) \rangle_{\mathbf r} + \frac{N-1}{N}\langle J_n^2 (\mathbf r) \rangle_{\mathbf r}^2 \right] -2 \left[\frac{1}{N^2}\langle J_n ^{4}(\mathbf r) \rangle_{\mathbf r} + \frac{N-1}{N^2}\langle J_n^2 (\mathbf r) \rangle_{\mathbf r}^2 \right] + \left[\frac{1}{N^3}\langle J_n^{4}(\mathbf r)\rangle_{\mathbf r} + \frac{3(N-1)}{N^3}\langle J_n^{2}(\mathbf r)\rangle_{\mathbf r}^{2}\right]\\
		&= \frac{(N-1)^2}{N^3}\langle J_n^{4}(\mathbf r)\rangle_{\mathbf r} + \frac{(N-1)\left((N-1)^2+2\right)}{N^3}\langle J_n^{2}(\mathbf r)\rangle_{\mathbf r}^{2}
	\end{align}
	Putting the above together, we can calculate $\mathrm{Var}_{\mathbf r}(X)$ as follows:
	\begin{align}
		\mathrm{Var}_{\mathbf r}(X) &= \langle X^2\rangle_{\mathbf r} - \langle X\rangle_{\mathbf r}^2 \\
		&= \frac{(N-1)^2}{N^3}\langle J_n^{4}(\mathbf r)\rangle_{\mathbf r} + \frac{(N-1)\left((N-1)^2+2\right)}{N^3}\langle J_n^{2}(\mathbf r)\rangle_{\mathbf r}^{2} - \left(\frac{N-1}{N}\langle J_n ^{2}(\mathbf r) \rangle_{\mathbf r}\right)^2 \\
		&= \frac{(N-1)^2}{N^3}\left( \langle J_n^{4}(\mathbf r)\rangle_{\mathbf r} -\langle J_n ^{2}(\mathbf r) \rangle_{\mathbf r}^2\right) + \frac{2(N-1)}{N^3}\langle J_n ^{2}(\mathbf r) \rangle_{\mathbf r}^2 \\
		&= \frac{(N-1)^2}{N^3} \mathrm{Var}_{\mathbf r}(J_n^2(\mathbf r)) + \frac{2(N-1)}{N^3}
	\end{align}
	Finally, dividing it by $\langle X\rangle_{\mathbf r}^2 = \frac{(N-1)^2}{N^2}$ provides Eq. (5) in the main text.

	\begin{figure*}[b]
		\centering
		\includegraphics[width=0.85\textwidth,
		trim=205pt 2pt 80pt 2pt,clip]{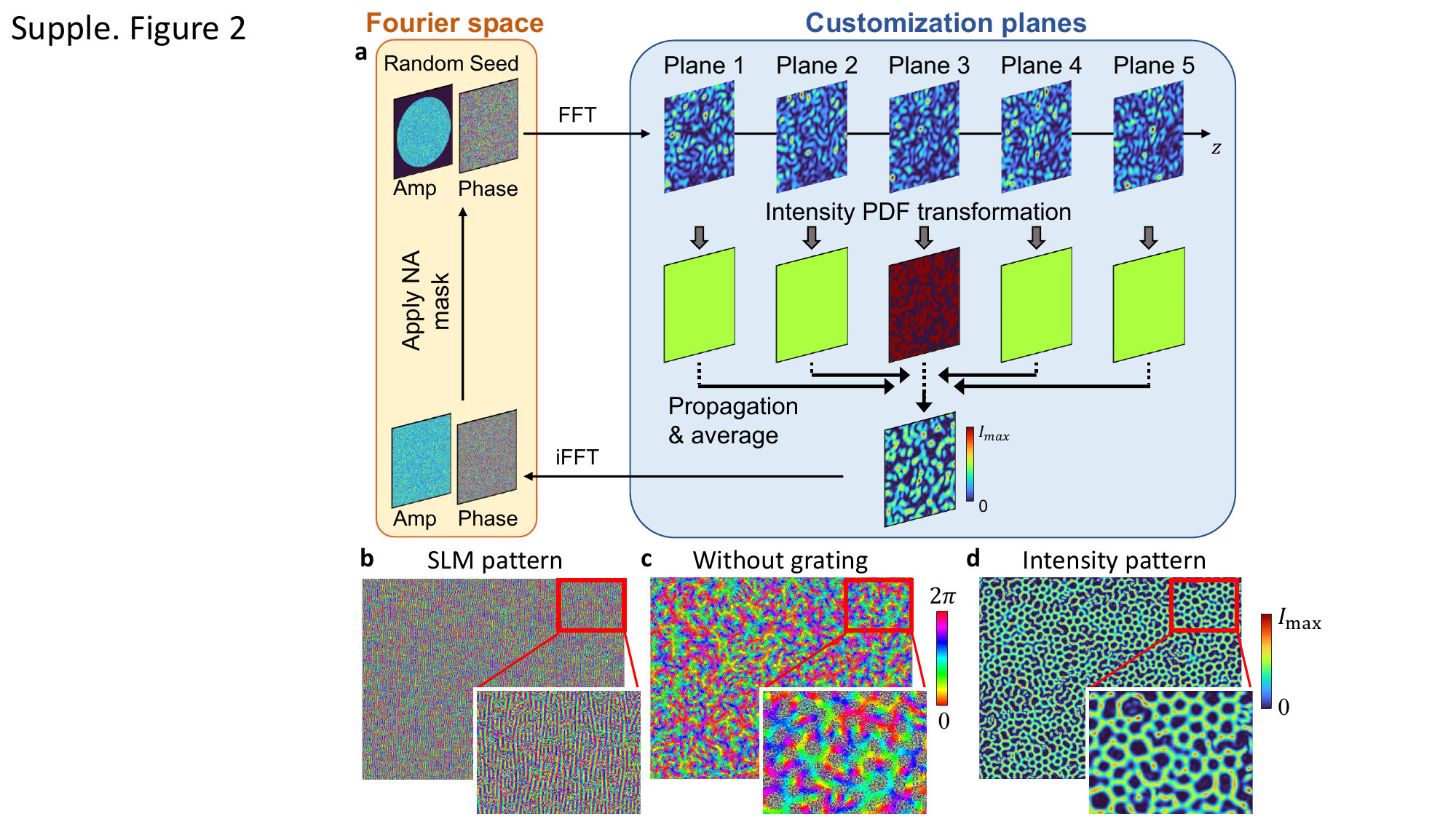} 
		\caption{\textbf{Numerical generation of customized speckles.} (a) Algorithm for tailoring speckle intensity statistics in five axial planes at $z = 0, \pm 0.7 z_R, \pm 1.4 z_R$. We apply an NA mask to fields in Fourier space and impose the target intensity PDF in each plane. Starting from a random seed, the optimization process iterates until convergence. (b) Example SLM phase pattern used for experimental generation of 3D tailored speckles in Fig.~1 of the main text. (c) SLM phase modulation pattern optimized for the same target field as in (b), but with the blazed grating removed for visualization. (d) An example of a binary speckle pattern at the focal plane $z=0$ produced by the SLM pattern in (b). The FOV contains $1.5\times10^4$ speckle grains.}
		\label{fig_supp_algorithm}
	\end{figure*}
	
	\section{Tailored speckle generation}
	
	Generation of tailored 3D speckle fields requires spatial modulation of both the amplitude and phase of a monochromatic light. In our previous work of tailoring 3D speckle statistics \cite{han2023tailoring}, a phase-only SLM is placed at the Fourier plane of a sample. The phase pattern of SLM macro-pixels is directly optimized using the measured transmission matrix (TM) that maps the field distribution from the SLM plane to the sample plane. However, the dimension of TM scales with the number of speckle grains in the field of view (FoV) and becomes excessively large for a broad FoV.
			
	Here, we move the phase-only SLM to the conjugate plane of the sample at $z=0$. To generate customized fields with a phase-only SLM, we use a two-step optimization procedure: (1) to obtain the target intensity probability density function (PDF) in all axial planes by optimizing the complex field distribution in the focal plane ($z=0$), and (2) to realize the optimized field from (1) using phase-only modulation of SLM. Each step is described in the following subsections. All code is available on \href{https://github.com/SeungYunH/tailored-speckle-imaging}{Github}.
	
	\subsection{Speckle field optimization} \label{sec:supp_field_optimizations}
	
	We optimize a complex field distribution in the sample plane ($z=0$) such that, after propagation in free space, the intensities in each designated axial plane satisfy the prescribed PDF (Fig.~\ref{fig_supp_algorithm}a). The algorithm proceeds as follows. A random complex field is initialized in the Fourier domain within the numerical aperture (NA) support, then transformed via Fast Fourier Transform (FFT) to real space at the focal plane ($z=0$, Plane 3 in Fig.~\ref{fig_supp_algorithm}a). The field is numerically propagated to four axial planes at $z = \pm 0.7 z_R, \pm 1.4 z_R$ (Planes 4, 3, 5, 1) using the beam propagation method, where $z_R$ denotes the full width at half maximum (FWHM) of axial intensity decorrelation of Rayleigh speckles. In each plane, the intensity values are transformed to satisfy the target intensity PDF, known as ``intensity PDF transformation'' \cite{bender2019creating}, while the field phase pattern is preserved. Since the intensity transformation breaks the field propagation relation between different planes, the modified field patterns at $z = \pm 0.7 z_R, \pm 1.4 z_R$ are propagated back to $z=0$ and averaged together with that at $z=0$. An inverse FFT (iFFT) then returns the field to the Fourier domain, where the NA mask is applied to retain only the allowed spatial frequencies. The above procedure is iterated to optimize a single NA-limited field pattern at $z = 0$ so that the resulting intensity PDFs at five axial planes ($z = 0, \pm 0.7 z_R, \pm 1.4 z_R$) approach the target PDFs. Here, we track the optimization error by the $L_1$ norm of the difference between the obtained and target PDFs. We run the algorithm for $10^4$ iterations, at which point the relative decrease of the optimization error per iteration has saturated to about $10^{-7}$. 
	
	To account for experimental imperfections in the optics, we restrict the NA support to 80\% of the objective NA (an effective $\text{NA}=0.95\times0.8=0.76$), which prevents the loss of spatial frequency content near the maximum. For consistency, the Rayleigh speckles used for comparison are generated with the same 80\% NA.
	
	Unlike our previous method \cite{han2023tailoring}, we assume an ideal Fourier transform by the lens and a precise alignment of the optics, ignoring optical aberrations and misalignments that have been included in the previously measured TM. Experimentally, upon refining the optical alignment, the measured intensity pattern has 94\% correlation with the target pattern of tailored speckle statistics. Furthermore, the use of FFT instead of TM multiplication significantly boosts the optimization speed.
	
	\vspace{-10pt}
	
	\subsection{SLM pattern optimization} \label{sec:supp_SLM_optimizations}
	
	The optimized complex field distribution in the sample plane ($z=0$) is realized by a phase-only SLM in a conjugate plane. The SLM is uniformly illuminated by a monochromatic laser beam at $\lambda$ = 488 nm. Ignoring lens aberration and misalignment of the relay optics, we numerically optimize the SLM phase pattern that generates the target field distribution at $z=0$ through first-order diffraction filtering. The propagation of light from the SLM to $z=0$ is simulated with a differentiable forward model to enable gradient-based optimization. Let $\phi(\mathbf{r})$ denote the SLM phase pattern. The field immediately after the SLM is
	\begin{equation}
		E_{\mathrm{SLM}}(\mathbf{r}) = \exp\!\big[i\phi(\mathbf{r})\big].
	\end{equation}
	The field delivered to the sample is then obtained by retaining the first diffraction order in the Fourier domain,
	\begin{equation}
		\hat{y} = \mathcal{F}^{-1}\!\left\{ H_{1}(\mathbf{k})\,\mathcal{F}\!\left[ 
		E_{\mathrm{SLM}}(\mathbf{r}) \right] \right\},
	\end{equation}
	where $\mathcal{F}[\cdot]$ is the Fourier transform, $H_{1}(\mathbf{k})$ is a bandpass filter centered on the first diffraction order, and $\hat{y}$ is the predicted complex field in the sample plane.
	
	We optimize $\phi(\mathbf{r})$ by minimizing the mismatch between $\hat{y}$ and the target field $E_{\mathrm{target}}$, with the loss function
	\begin{equation}
		\mathcal{L}(\phi)
		= \left\lVert \hat{y} - E_{\mathrm{target}} \right\rVert_2^2
		- \alpha \left\lVert \hat{y} \right\rVert_2^2,
	\end{equation}
	where the first term enforces field matching and the second term penalizes a low-power solution. We set $\alpha = 0.2$, chosen empirically to balance field matching and power loss.
	
	We solve this optimization using Nesterov accelerated gradient (NAG) for 100 iterations. The resulting phase patterns generate the complex field distribution in the sample plane with Pearson correlations of $82\%$--$99\%$ relative to the target $E_{\mathrm{target}}$, while retaining $58\%$--$68\%$ of power. The lower correlation and power values correspond to Rayleigh and tailored speckles, whereas the higher values correspond to periodic fringes. An example SLM phase pattern is shown in Fig.~\ref{fig_supp_algorithm}b. Fig.~\ref{fig_supp_algorithm}c shows the same pattern with the grating removed, revealing the underlying phase structure. The corresponding output intensity pattern $|\hat{y}|^{2}$ is shown in Fig.~\ref{fig_supp_algorithm}d. In this example pattern, $1.5\times10^4$ speckle grains are generated using $600\times 792 = 4.8\times10^5$ SLM pixels.

	\begin{figure*}[h!]
		\centering
		\includegraphics[width=0.85\textwidth,
		trim=120pt 140pt 140pt 122pt,clip]{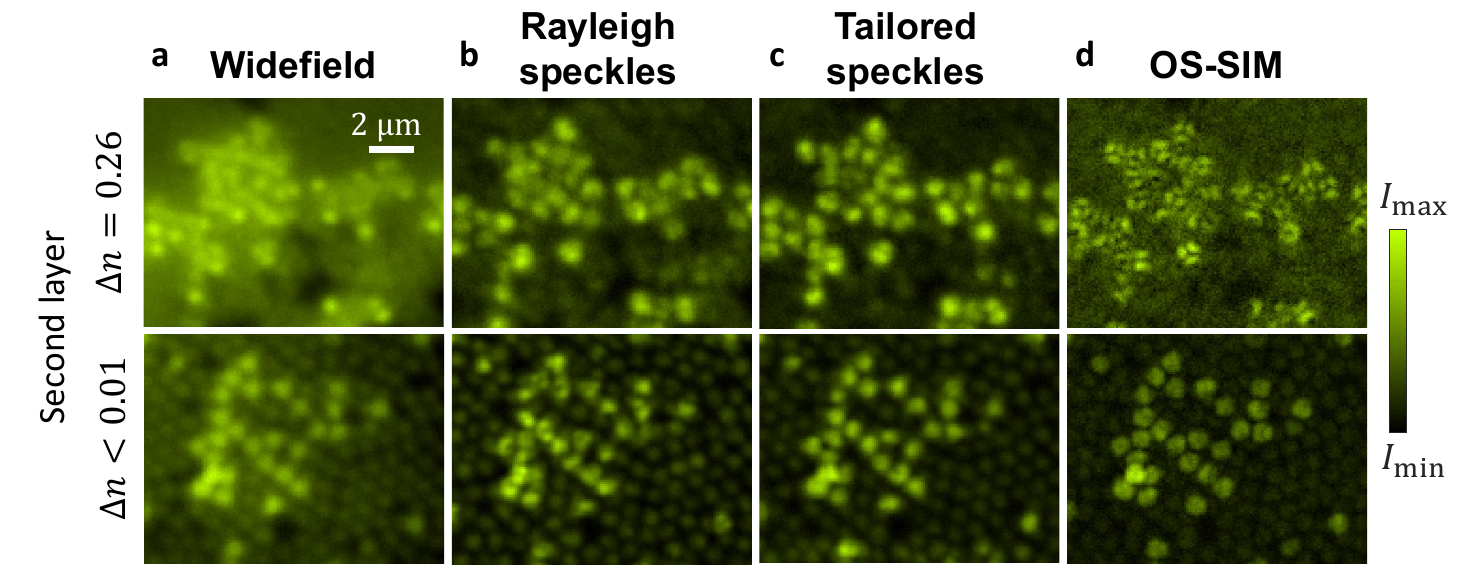} 
		\caption{\textbf{Imaging of second-layer fluorescent beads in index-matching liquid.}
			Corresponding to Fig.~3 in the main text, reconstructed images of fluorescent beads in the second layer in index-mismatched water (top row) and index-matched liquid (bottom row), obtained with widefield imaging (a), DSI with Rayleigh speckles (b) and tailored speckles (c), and OS-SIM (d). Image artifacts with OS-SIM in the index mismatch case are mostly removed once the refractive indices are matched.}\label{fig_supp_RImatchBeads}
	\end{figure*}
	\section{Reduction of sample scattering} \label{sec:supp_Index_match}
	
	Due to refractive index mismatch between fluorescent beads and surrounding water ($\Delta n = 0.26 $), the second-layer beads are poorly reconstructed by OS-SIM, as shown in Fig.~3 of the main text. To verify that this degradation is caused by sample-induced scattering and aberration, we repeat the experiment after replacing water by a refractive-index-matching liquid ($n_m \approx 1.59$, Cargille) to the polystyrene bead ($\Delta n < 0.01$).
	
	As shown in the bottom row of Fig.~\ref{fig_supp_RImatchBeads}, OS-SIM largely recovers image quality when scattering is minimized. This confirms that the artifacts in OS-SIM reconstruction with index mismatch are primarily due to sample-induced aberration and scattering.

	
	\section{Incoherent OS-SIM} \label{sec:supp_incSIM} 

	\begin{figure*}[b]
		\centering
		\includegraphics[width=0.9\textwidth,
		trim=45pt 90pt 90pt 97pt,clip]{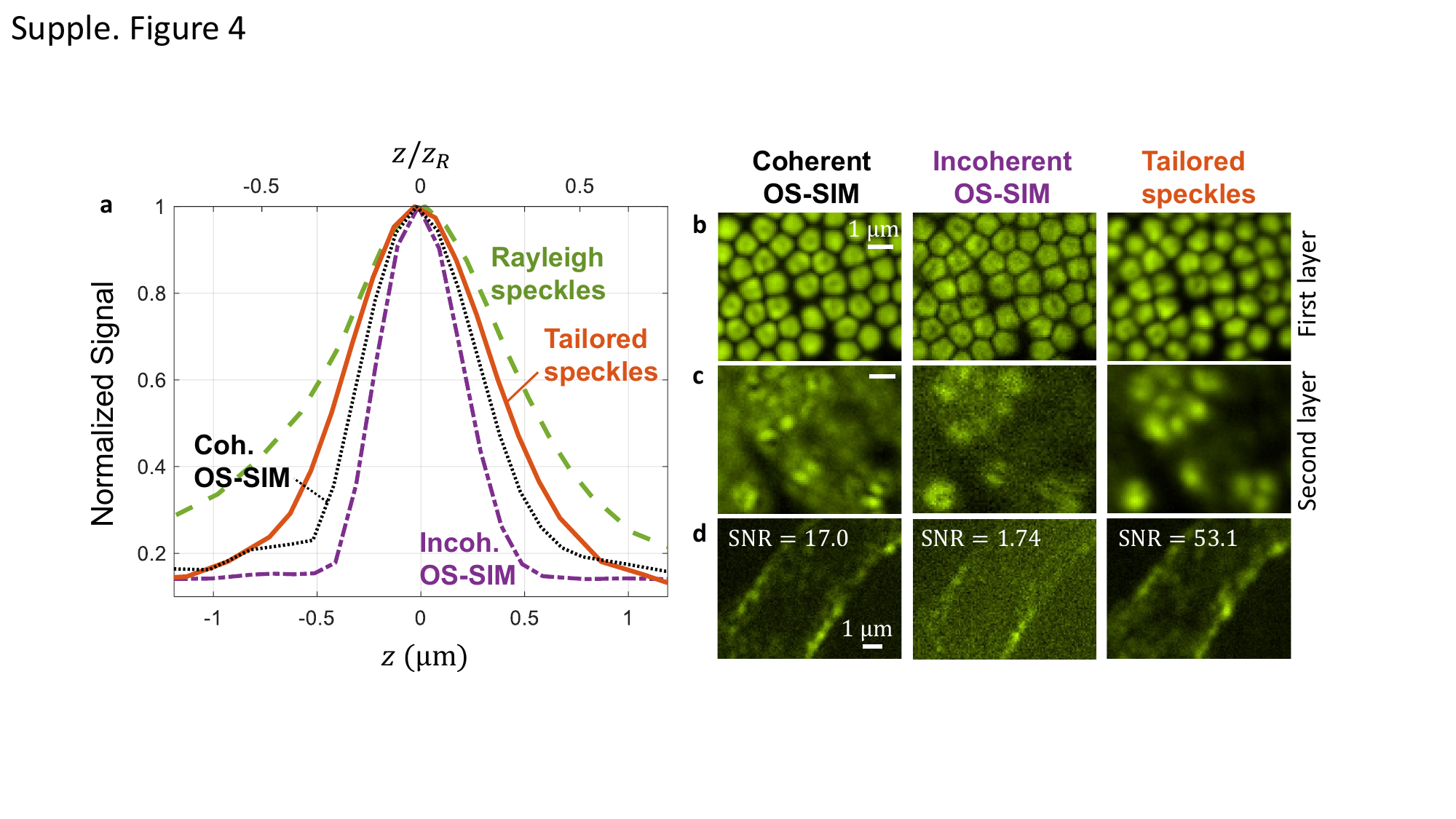} 
		\caption{\textbf{Incoherent OS-SIM.}
			(a) Normalized signal of a thin fluorescent film as a function of axial position, for comparison of coherent and incoherent OS-SIM, DSI with Rayleigh speckles and tailored speckles.
			(b,c) Reconstructed images of stacked fluorescent beads in the first and second layers (immersed in water), obtained with coherent and incoherent OS-SIM and DSI with tailored speckles. Incoherent OS-SIM suffers more image artifacts and noise due to the loss of fringe contrast by light scattering.
			(d) Reconstructed images of a mouse brain blood vessel with coherent and incoherent OS-SIM and DSI with tailored speckles. Incoherent OS-SIM yields the lowest SNR, and DSI with tailored speckle illumination achieves the highest SNR.
		}\label{fig_supp_incSIM}
	\end{figure*}
    
	In this section, we compare the incoherent OS-SIM to DSI with tailored speckles. Unlike coherent OS-SIM, incoherent OS-SIM uses intensity fringes generated by fields with random phase fluctuations that are spatially uncorrelated and faster than the detector integration time. Experimentally, we emulate incoherent illumination by introducing controlled phase fluctuations to the fringe pattern. A coherent fringe pattern is generated by a grating in the SLM, $E(x) = (e^{ikx} + e^{-ikx})e^{ik'x}$, where $e^{ik'x}$ is a position-dependent phase that does not affect the intensity pattern $I_s(x) = |E(x)|^2$ that is mapped to the sample plane. Averaging the fringe intensities over different $k'$ values approximates an ensemble average over random phases. In our implementation, we select nine $k'$ values, uniformly spaced within the NA of the imaging optics, to illuminate the sample. The nine corresponding fluorescence images are averaged to form a single image for incoherent OS-SIM. Image reconstruction of incoherent OS-SIM then follows the same procedure as coherent OS-SIM (main text). A further increase in the number of $k'$s beyond nine does not show a significant impact on the reconstructed image quality.  
	
	Incoherent illumination in OS-SIM has stronger optical sectioning than coherent OS-SIM, as shown in Fig.~\ref{fig_supp_incSIM}a. This is because, in out-of-focus planes, fringes corresponding to different $k'$ values are laterally shifted by different amounts, and averaging the intensities over different $k'$s washes out the fringe contrast and suppresses the OS-SIM signal from those planes. This selective suppression of out-of-focus signal enhances axial sectioning relative to coherent illumination.

    However, incoherent illumination results in lower fringe contrast than coherent illumination even when in focus. Effectively incoherent intensity fringes after the SLM, $I_s(x) = |E(x)|^2 = 2+e^{2ikx} + e^{-2ikx}$, are mapped by the optical transfer function (OTF) of the imaging system to the sample plane $z=0$. The OTF is peaked at zero spatial frequency (DC) and decreases at higher frequencies. $\pm2k$ components in $I_s(x)$ are attenuated relative to the DC term, lowering the fringe contrast in the sample $z=0$. Because the incoherent OS-SIM produces a weaker signal, it is more susceptible to sample-induced aberration and scattering, as shown below.  
	
	Experimentally, we compare incoherent OS-SIM to coherent OS-SIM and DSI with tailored speckle illumination of fluorescent beads in water (refractive-index mismatch $\Delta n = 0.26$). For a fair comparison, each frame of coherent OS-SIM and DSI is averaged over nine repeated measurements to match the noise level of incoherent OS-SIM. In the reconstructed images of the first layer (Fig.~\ref{fig_supp_incSIM}b), incoherent OS-SIM shows more image artifacts: less uniform intensity distribution within individual beads, and polygonal rather than circular boundaries of beads. The second layer image (Fig.~\ref{fig_supp_incSIM}c) shows more severe degradation for incoherent OS-SIM, due to the reduced fringe contrast compared to coherent OS-SIM. DSI with tailored speckles outperform both coherent and incoherent OS-SIM, displaying clearer bead boundaries and a more suppressed background. 
	
	When imaging the mouse brain blood vessel (Fig.~\ref{fig_supp_incSIM}d), incoherent OS-SIM yields the lowest SNR of 1.74, while coherent OS-SIM has a notably higher SNR of 17.0. DSI with tailored speckles achieves the highest SNR of 53.1 with the best visibility of the vascular structure. The SNR values here are higher than in Fig.~4 of the main text because each frame is an average of nine measurements. Together, these results show that incoherent illumination in OS-SIM improves axial sectioning but are more susceptible to sample-induced scattering, whereas DSI with tailored speckle illumination remains robust to strong scattering.
	

	\section{Spatial power spectrum of tailored speckles } \label{sec:supp_powerSpectrum} 

	\begin{figure*}[b]
		\centering
		\includegraphics[width=0.98\textwidth,
		trim=10pt 125pt 20pt 100pt,clip]{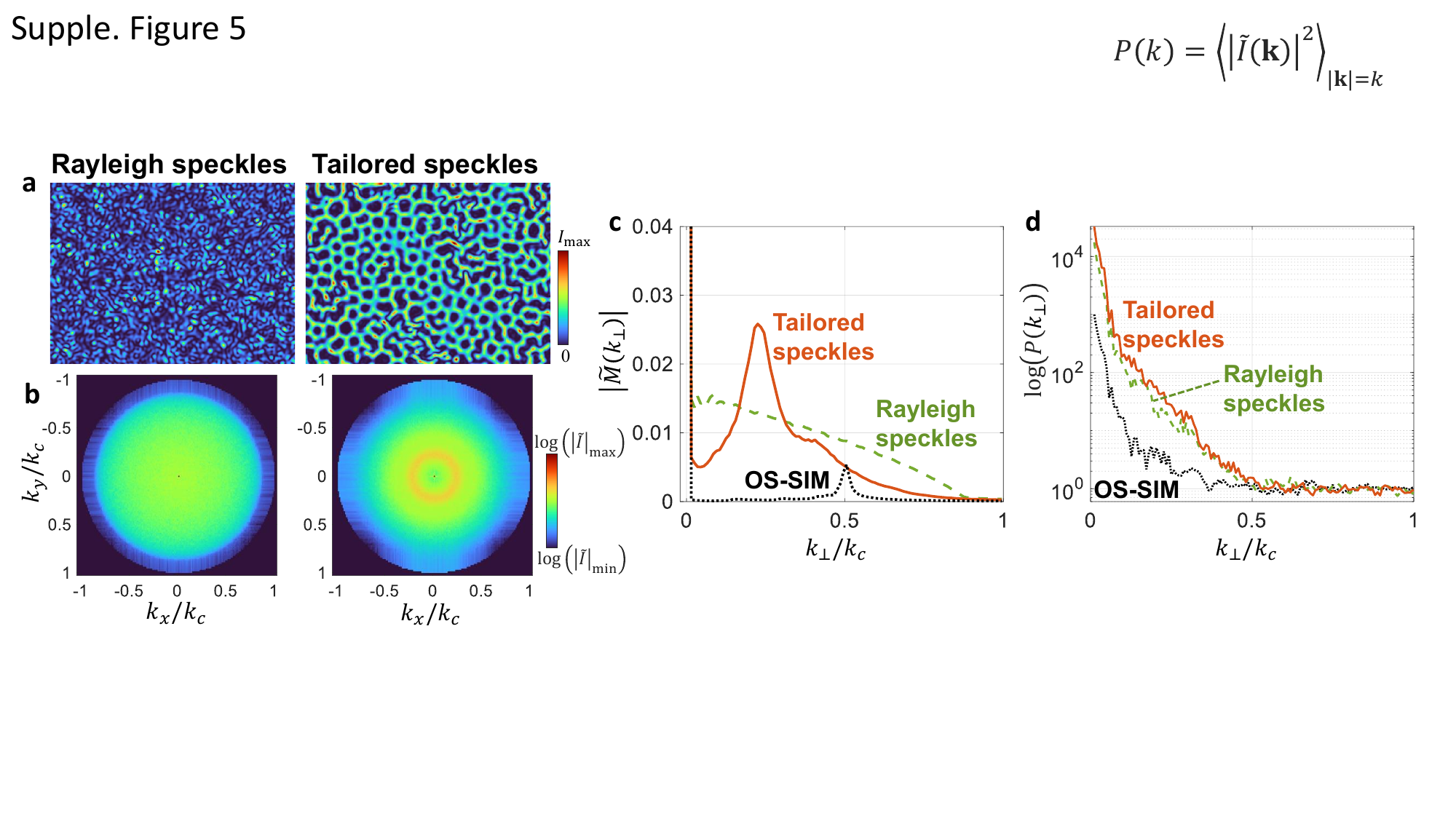} 
		\caption{\textbf{Spatial power spectra of Rayleigh and tailored speckles.}
			(a) Representative in-focus illumination intensity patterns for Rayleigh speckles (left) and tailored speckles (right). The in-focus tailored speckle pattern has binary intensity values. 
			(b) 2D spatial power spectra, $|\tilde{M}(\mathbf{k}_\perp)|$, averaged over 30 independent illumination patterns and plotted on a logarithmic scale. The excitation patterns use up to 80\% of the objective NA to avoid loss of spatial frequency content near the cutoff. The axes are normalized by the detection cutoff frequency $k_c$.
			(c) Azimuthal average of spatial power spectra in (b) as a function of transverse spatial frequency $k_\perp$, normalized by the detection cutoff frequency $k_c$ for tailored speckles (orange solid), Rayleigh speckles (green dashed) and sinusoidal fringes for OS-SIM (black dotted). Binary speckle pattern in (a) has weaker high-frequency content than the Rayleigh speckle.
			(d) Azimuthal-averaged spatial power spectra $P(k_\perp)$ for reconstructed images of mouse brain blood vessels in Fig.~4c of the main text, normalized by the noise level and plotted on a logarithmic scale, obtained with DSI with tailored speckles (orange solid), Rayleigh speckles (green dashed), and sinusoidal fringes for OS-SIM (black dotted). Despite weaker high-frequency illumination content, tailored speckles yield stronger high-spatial-frequency reconstruction than Rayleigh speckles. This indicates that the enhancement is not due to increased high-spatial-frequency components in illumination patterns but perhaps related to the smaller reduction in spatial contrast of fluorescence images after convolution with the detection PSF for tailored speckles.
		}\label{fig_supp_powerSpectrum}
	\end{figure*}

	To quantify how speckle customization redistributes spatial frequencies, we compare the spatial power spectrum of tailored speckles to that of Rayleigh speckles (Fig.~\ref{fig_supp_powerSpectrum}a). We perform a spatial Fourier transform of the illumination intensity pattern at $z=0$, $\tilde{M}(\mathbf{k})=\mathcal{F}\{M(\mathbf{r})\}$ and display its magnitude and azimuthal average in Figs.~\ref{fig_supp_powerSpectrum}b and c for Rayleigh and tailored speckles, respectively. Here, $\tilde M(\mathbf{k})$ is normalized to have the same DC magnitude $\left|\tilde M(\mathbf{0})\right|=1$, and the spatial frequencies $k_x$, $k_y$, and $k_{\perp} = (k_x^2+k_y^2)^{1/2}$ in x, y, and transverse directions, respectively, are normalized by the cutoff frequency $k_c$ of the detection PSF. Note that excitation patterns use up to 80\% of the objective NA to avoid loss of spatial frequency content near the cutoff. Compared to Rayleigh speckles, tailored speckles have more frequencies near $0.23\,k_c$ and less above $0.31\,k_c$ in Fig.~\ref{fig_supp_powerSpectrum}c. 
	
	Fig.~4c of the main text shows the spatial power spectra for reconstructed images of the mouse brain blood vessels. In Fig.~\ref{fig_supp_powerSpectrum}d, the azimuthal average of the spatial power spectra, normalized by the noise level, is plotted on a logarithmic scale. The result reveals that for this sample, the tailored speckle illumination yields slightly stronger high-spatial-frequency components than Rayleigh speckles, including frequencies beyond $0.31\,k_c$, despite the weaker high-frequency content of the illumination pattern itself. This suggests that tailored speckle illumination preserves the spatial resolution in the reconstructed image. For comparison, Figs.~\ref{fig_supp_powerSpectrum}c and d also include the normalized azimuthal average for OS-SIM, whose power spectrum of the reconstructed image is notably lower than those of Rayleigh and tailored speckles. 

    \vspace{20pt}
	\section{Full Field-of-view imaging} \label{sec:supp_FullFOV} 
\begin{figure*}[htbp]
		\centering
		\includegraphics[width=0.85\textwidth,
		trim=110pt 175pt 110pt 150pt,clip]{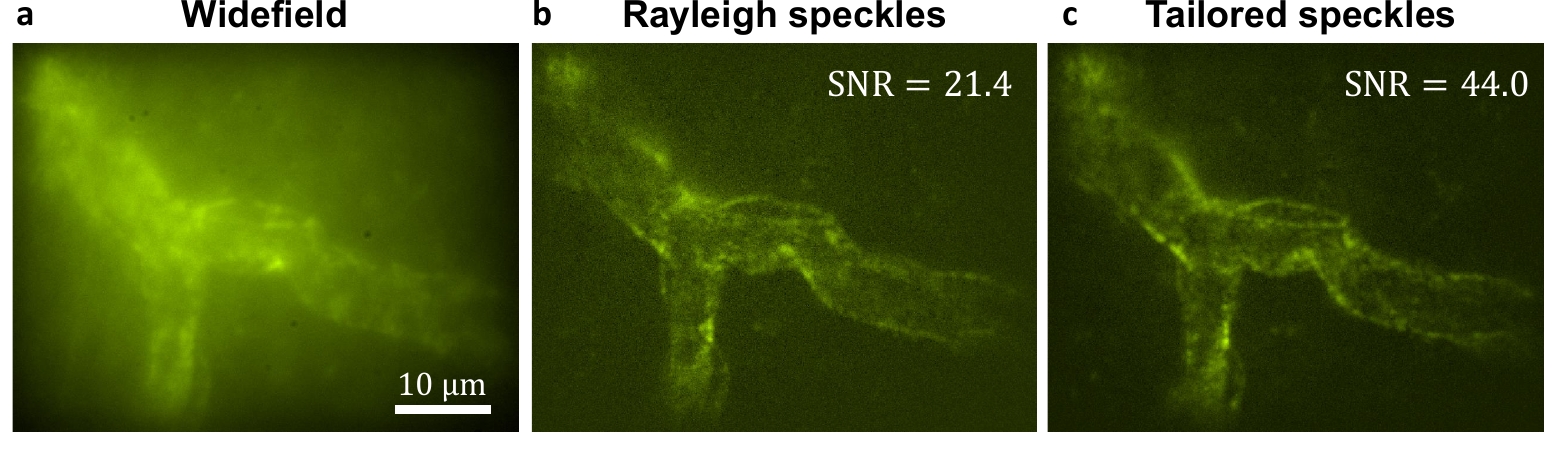} 
		\caption{\textbf{Full FOV imaging with tailored speckle illumination.}
			Reconstructed images of mouse brain blood vessels with widefield imaging (a), DSI with Rayleigh speckles (b) and tailored speckles optimized for full FOV (c). The SNR with tailored speckles is $2\times$ higher than with Rayleigh speckles, similar to Fig. 4 in the main text.
		}\label{fig_supp_FullFOV}
	\end{figure*}

	While we tailored speckle statistics within a reduced FOV in the main text, full-FOV tailored speckles provide a similar imaging quality. We optimize $1.5\times10^{4}$ speckle grains in total, for $\times4$ larger area than in the main text. Fig.~\ref{fig_supp_FullFOV}a-c shows reconstructed images of the mouse brain blood vessels at a depth of 20~\textmu m using widefield imaging, DSI with Rayleigh speckles and tailored speckles, respectively, for the full FOV. DSI with tailored speckles provides stronger axial sectioning and $2\times$ higher SNR than with Rayleigh speckles, similarly to Fig. 4 in the main text.

    \bibliography{main}